\documentclass[reprint, amsmath, amssymb, aps, prd, superscriptaddress, nofootinbib, nobibnotes, longbibliography, floatfix, colorlinks=true, linkcolor=red, citecolor=blue]{revtex4-2}
\usepackage[utf8]{inputenc}
\usepackage{graphicx}
\graphicspath{{figures/}}
\usepackage{subfig}

\usepackage{bm}
\usepackage{multirow}

\usepackage{amsmath}
\usepackage[colorinlistoftodos]{todonotes}
\usepackage[colorlinks=true, allcolors=blue]{hyperref}
\usepackage{acro}

\DeclareAcronym{GR}{
	short = GR,
	long  = general relativity
	}
	
\DeclareAcronym{BH}{
	short = BH ,
	long  = black hole
}

\DeclareAcronym{BBH}{
	short = BBH ,
	long  = binary black hole
}

\DeclareAcronym{GW}{
	short = GW ,
	long  = gravitational wave
}

\DeclareAcronym{CBC}{
	short = CBC,
	long  = compact binary coalescence
}

\DeclareAcronym{SNR}{
	short = SNR,
	long  = signal-to-noise ratio
}

\DeclareAcronym{QNM}{
	short = QNM,
	long  = quasi-normal mode
}

\DeclareAcronym{QQNM}{
	short = QQNM,
	long  = quadratic quasi-normal mode
}

\usepackage{orcidlink}
\usepackage[capitalise]{cleveref}
\crefname{figure}{Fig.}{Figs.}
\Crefname{figure}{Fig.}{Figs.}

\newcommand{\reply}[1]{{\color{black}{#1}}}

\newcommand{\bayesf}{{\color{black}{62}}}
\newcommand{\starttime}{{\color{black}{4 $M_\mathrm{f}$}}}
\newcommand{\quasigma}{\textcolor{black}{3.4~$\sigma$}}

\newcommand{\MismatchTenFifteenModes}{\textcolor{black}{0.69\%}}

\newcommand{\expectedsnr}{\textcolor{black}{2.2}}
\newcommand{\SNROneSigma}{\textcolor{black}{$[1.5, 3.4]$}}

\begin{document}
\title{A nonlinear voice from GW250114 ringdown}
\author{Yi-Fan Wang \orcidlink{0000-0002-2928-2916}}
\affiliation{Max-Planck-Institut f{\"u}r Gravitationsphysik (Albert-Einstein-Institut), Am M{\"u}hlenberg 1, D-14476 Potsdam, Germany}
\affiliation{Purple Mountain Observatory, Chinese Academy of Sciences, Nanjing 210034, China}

\author{Sizheng Ma \orcidlink{0000-0002-4645-453X}}
\affiliation{Perimeter Institute for Theoretical Physics, Waterloo, ON N2L2Y5, Canada}

\author{Neev Khera \orcidlink{0000-0003-3515-2859}}
\affiliation{Department of Astronomy, Tsinghua University, Beijing 100084, China}

\author{Junquan Su \orcidlink{0009-0008-1901-533X}}
\affiliation{Department of Astronomy, Tsinghua University, Beijing 100084, China}

\author{Huan Yang \orcidlink{0000-0002-9965-3030}}
\email{hyangdoa@tsinghua.edu.cn}
\affiliation{Department of Astronomy, Tsinghua University, Beijing 100084, China}

\begin{abstract}    
Gravitational-wave astronomy, by detecting ripples in spacetime, has opened a new window to observe compact objects and probe theories of gravity in the nonlinear strong-field regime. The ringdown signal of a binary black hole merger contains a superposition of damped sinusoids known as quasi-normal modes \cite{qnm1}, whose frequencies are completely determined by the mass and spin of the remnant black hole and form the basis of \textit{black hole spectroscopy} \cite{Nohairseminal1, Nohairseminal2, Nohairseminal3}. A crucial prediction yet to be observationally confirmed is the existence of quadratic quasi-normal modes, which represent fundamental properties associated with wave–wave coupling in general relativity \reply{, and the leading mode} is predicted to be detectable with next-generation ground-based detectors \cite{Mitman:2022qdl, Cheung:2022rbm, Khera:2024bjs} \reply{using traditional methods}. Here we show the first observational evidence for a set of quadratic quasi-normal modes in the ringdown of the binary black hole merger GW250114, the loudest gravitational-wave event detected to date\reply{, enabled by a novel analysis}. These nonlinear modes result from the quadratic coupling of the linear $(2,2,n)$ modes with $n\leq3$. Starting the analysis at \reply{a time corresponding to four times the} remnant mass ($M_\mathrm{f}$) after the merger, the evidence for their presence reaches a Bayes factor of \bayesf{}. A phenomenological test \reply{allowing these modes to deviate from the theoretical prediction} rejects the zero-amplitude hypothesis at a significance of \quasigma{}, while \reply{the inferred amplitude and complex frequency are consistent with the prediction of general relativity}. This finding provides the first observational evidence of gravitational wave-wave interaction and extends black hole spectroscopy from the linear to the nonlinear regime. It also establishes a new direction for testing the fundamental nonlinear structure of general relativity with the most extreme gravity.
\end{abstract}

\maketitle
\acresetall

\section{Introduction} \label{sec:intro}

The \acp{GW} emission during the \acl{BH} ringdown stage contains a superposition of damped sinusoidal signals, known as \acp{QNM} \cite{qnm1}.
According to the no-hair theorem \cite{Nohairseminal1, Nohairseminal2, Nohairseminal3}, the frequencies and damping times of these modes are uniquely determined by the mass and spin of the remnant \acl{BH}.
Detecting multiple \acp{QNM} thus enables independent inference of the \acl{BH}'s mass and spin and verification of the no-hair theorem.
This approach, often termed black hole spectroscopy \cite{Dreyer:2003bv,Berti:2025hly}, offers a means to test \ac{GR} in the strong-field regime surrounding a Kerr \acl{BH}.
Beyond the linear \ac{QNM} spectrum, the ringdown also contains non-modal signals \cite{Lu:2025vol,Oshita:2025qmn}, tail \cite{Ma:2024hzq,DeAmicis:2024eoy,Islam:2024vro,DeAmicis:2024not}, dynamical excitation, and nonlinearities.
For example, quadratic modes \cite{Mitman:2022qdl, Cheung:2022rbm, Ma:2024qcv,Khera:2024bjs,Khera:2023oyf} arise from second-order perturbation and provide a direct probe into the nonlinear dynamics of \ac{GR}. 
Detecting such quadratic modes is thus crucial for understanding the full theoretical structure of black hole perturbation. 
\reply{In addition, as they are sourced by the oscillating piece of the stress-energy of linear waves, their detection is also an observational proof for ``gravitational waves gravitate''.}

Extracting the rich \ac{QNM} spectrum in the ringdown signals poses data analysis challenges, partly due to the need to isolate the ringdown stage from the preceding inspiral and merger stages.
A variety of strategies have been developed to address this issue, including time- or frequency-domain methods \cite{Isi:2019aib,Carullo:2019flw, Capano:2021etf,Isi:2021iql,Wang:2023ljx,Ma:2022wpv,Ma:2023vvr,Ma:2023cwe,Finch:2021qph,Brito:2018rfr, Ghosh:2021mrv, Maggio:2022hre,Giesler:2024hcr,Pompili:2025cdc, Dong:2025igh, Wang:2024jlz,Chandra:2025ipu}.
Historically, the first observational evidence of a fundamental $(\ell,m,n) = (2,2,0)$ \ac{QNM}, where $\ell,m$ are the polar and azimuthal numbers and $n$ denotes the overtone number, came from the first detection in GW150914 \cite{LIGOScientific:2016lio}, and has since been further probed in subsequent events \cite{LIGOScientific:2019fpa,LIGOScientific:2020tif,LIGOScientific:2021sio}.
Evidence of a subdominant fundamental mode $(3,3,0)$ was reported by Ref.~\cite{Capano:2021etf} from GW190521 \cite{LIGOScientific:2020iuh,LIGOScientific:2020ufj}, attributed to the large remnant mass and asymmetric mass ratio; an alternative interpretation accounting for (2,1,0) modes in the presence of precessing spins is discussed in \cite{Siegel:2023lxl}.
More recently, GW231123 exhibited evidence of modes beyond the dominant (2,2,0) \ac{QNM} \cite{LIGOScientific:2025rsn, Siegel:2025xgb}, but its complex source properties established difficulty on mode identification.
Additionally,~\cite{Isi:2019aib} reported a detection of the overtone (2,2,1) in GW150914 by starting the analysis at the merger time.

However, starting the ringdown analysis close to the merger presents significant data analysis challenges given the current incomplete theoretical understanding of the involved components. This regime involves a complex mixture of \reply{fundamental QNMs, additional overtones} \cite{London:2014cma, Giesler:2024hcr}, non-modal signals (e.g., direct waves \cite{Lu:2025vol}), and nonlinearities such as quadratic modes \cite{Mitman:2022qdl, Cheung:2022rbm,Lagos:2022otp, Ma:2024qcv,Khera:2024bjs,Khera:2023oyf}. Without a theoretical framework, it is difficult to cleanly separate these contributions in a data-driven approach, especially when multiple modes are comparable in amplitude without clear hierarchical ordering. These challenges will become more pronounced in high \ac{SNR} events. Nevertheless, enabled by recent theoretical progress \cite{Khera:2023oyf}, we focus on detecting and isolating nonlinear quadratic modes by computing their amplitudes and separating this contribution from the overall signal.

Previous ringdown \reply{analyses} commonly excise the pre-ringdown data \cite{Carullo:2019flw,Capano:2021etf,Wang:2023ljx,Isi:2021iql,Ma:2023vvr,Ma:2022wpv,Ma:2023cwe} or model it in an agnostic \cite{Finch:2021qph} \reply{or parameterized effective-one-body waveform \cite{Brito:2018rfr,Ghosh:2021mrv,Maggio:2022hre,Pompili:2025cdc}}.
However, in principle, the amplitude and phase of each \ac{QNM} are determined by the pre-ringdown dynamics.
As evidence grows in \ac{GR}'s accuracy in describing the full waveform of binary black hole coalescence, the inspiral and merger phases should, in turn, provide informative constraints for ringdown analysis \cite{Pacilio:2024tdl,MaganaZertuche:2024ajz,Hughes:2019zmt,Lim:2022veo}. 
In this work, we introduce a novel methodology that integrates information from the pre-ringdown data analysis.
This is achieved by isolating the likelihood contributions of the inspiral-merger and ringdown stages, respectively, by gating and inpainting \cite{Zackay:2019kkv,Capano:2021etf,Wang:2023ljx, Correia:2023ipz} data outside the region of interest.
The posterior of the inspiral-merger phase is folded into the ringdown analysis as a \reply{physically informed} prior to boost the detectability.
This approach makes use of the inspiral-merger inference results while simultaneously ensuring an independent ringdown data analysis.

We apply this method to the black hole merger event GW250114, the loudest \ac{GW} detection to date with a network matched-filtering \ac{SNR} of 80 \cite{KAGRA:2025oiz, LIGOScientific:2025obp} reported by LIGO \cite{TheLIGOScientific:2014jea}, Virgo \cite{TheVirgo:2014hva} and KAGRA \cite{Aso:2013eba} collaboration.
The source is consistent with a binary of nearly equal-mass with small spins.
Previous analyses \cite{KAGRA:2025oiz, LIGOScientific:2025obp} have reported evidence for the fundamental \ac{QNM} $(2,2,0)$ and its first overtone $(2,2,1)$, \reply{and constrained the frequency of the fundamental $(4,4,0)$ mode}.
Building on this, we extend the investigation to the quadratic \acp{QNM} \reply{arising from the coupling among the $(2,2,n)$ modes.}
The coupling ratio between quadratic and linear modes is calculated theoretically.
The amplitudes of the linear modes are determined by a \reply{weighted} least-squares fit against \texttt{NRSur7dq4} \cite{Varma:2019csw}, a numerical relativity surrogate waveform.
This approach ensures a unified set of parameters to describe both the inspiral-merger and ringdown stages, thus we do not introduce any additional free parameters in the ringdown inference.
By \reply{separating} the quadratic modes from \reply{the total ringdown signal}, we establish a procedure to quantify their detection significance within a set of other signals.
The corresponding Bayes factor reaches \bayesf{} when evaluated from \starttime{} after the merger.
We also perform tests introducing fractional deviations to the theoretical prediction of amplitude \reply{and complex frequency} of the quadratic \acp{QNM}, and find both deviations to be consistent with zero, and rejecting a zero amplitude of the quadratic \acp{QNM} with significance of \quasigma.

\section{Method}
\label{sec:method}

\reply{
Despite their fundamental importance, the detection of \ac{QQNM} has generally been expected to require third-generation (3G) detectors, owing to their relatively small amplitudes \cite{Mitman:2022qdl, Cheung:2022rbm, Khera:2024bjs, Yi:2024elj}. Moreover, an event suitable for detecting quadratic QNMs must be sufficiently loud, yet particularly at early times close to the merger, multiple components coexist and do not exhibit a clear hierarchy. In other words, probing \ac{QQNM} requires a framework that quantifies the detectability of a target signal in data containing both \emph{other physical signals} and noise. This data-analysis challenge has become practically relevant only with the recent detection of GW250114.

To enhance sensitivity to weak nonlinear signals, we adopt two complementary strategies. First, we use early starting times to retain as much of the \ac{QQNM} as possible. However, we also observe that the modes sourced by different parent overtones tend to show significant cancellation effects (see \cref{sec:app:anatomy}). To robustly characterize wave-wave couplings, we combine the dominant QQNMs and search for this bundled nonlinear component, rather than targeting a single quadratic mode. In particular, we systematically compute the amplitudes of a set of \reply{ten} nonlinear quadratic modes within the $(4,4)$ multipole, generated from the coupling of any two linear $(2,2,n)$ \acp{QNM}, where $n \in \{0,1,2,3\}$. At \starttime{} after merger, the ten-mode sum has converged to the collective quadratic contribution from the $(2,2,n)$ couplings (see \cref{sec:app:anatomy}). At \starttime{}, the nominal theoretical \ac{SNR} of the bundled quadratic modes for GW250114 is expected to be $\sim\expectedsnr$. Second, we use information inferred from the inspiral-merger stage of the signal to physically inform the ringdown analysis. Related data-informed strategies have been discussed previously in \cite{Brito:2018rfr,Ghosh:2021mrv,Maggio:2022hre,Pompili:2025cdc} for ringdown and/or consistency tests. Although the inspiral-merger information formally enters the ringdown analysis in a manner similar to a prior, it is not a purely hand-chosen prior; rather, it is constrained by the observed gravitational-wave data through the inspiral-merger analysis. Equivalently, our procedure can be viewed as a detection of QQNMs using inspiral-merger-ringdown data.

To formulate the detection of a target signal in the presence of residual physical signals and noise, we generalize the standard Bayesian model-selection framework for a single mode. The usual single-mode test compares a signal-plus-noise hypothesis against a noise-only hypothesis through a Bayes factor. Here, instead, we compare a model containing the target signal, residual signal, and noise against a model containing only the residual signal and noise:
\begin{align}
\mathcal{B}
&= \frac{
p(d \mid \mathrm{target\ signal}+\mathrm{residual\ signal}+\mathrm{noise})
}{
p(d \mid \mathrm{residual\ signal}+\mathrm{noise})
} \nonumber \\
&= \frac{
p(d \mid \mathrm{total\ signal}+\mathrm{noise})
}{
p(d \mid \mathrm{total\ signal}-\mathrm{target\ signal}+\mathrm{noise})
}.
\label{eq:bayesevidence}
\end{align}
This expression reduces to the standard single-mode Bayes factor when the residual signal is absent. 

Operationally, the Bayes evidence in \cref{eq:bayesevidence} can be evaluated in different ways. A ringdown-only test would require marginalization over the full parameter space of both the target and residual signals with unknown amplitudes, which typically demands a substantially higher ringdown SNR. In contrast, we derive the amplitudes of the parent modes from a weighted least-squares fit towards a numerical relativity surrogate waveform (\cref{app:least-square}), and using first-principle calculation from the second-order perturbation theory to obtain the quadratic modes' amplitudes (\cref{app:qqnm}). We also use inspiral-merger information to constrain the parameters of both the total signal and the target signal entering the evidence calculation. The resulting model comparison thus assumes the linear and nonlinear ringdown structure predicted by \ac{GR} and tests whether the data support the inclusion of quadratic modes. The full statistical framework is presented in \cref{sec:app:statistics}.

We also perform an independent test of the presence of \ac{QQNM} by allowing their amplitudes or frequencies to vary through common deviation parameters for all modes in the bundle, while keeping the residual ringdown signal fixed. The resulting amplitude posterior can be compared with the predictions of \ac{GR}. Similarly, the recovered frequencies can be compared with the expected \acp{QQNM} frequencies and with those of the linear \acp{QNM}, providing an additional check on the nonlinear interpretation of the detection.

We use numerical relativity simulations to validate the ten-mode truncation and the adopted starting time \starttime{}. Ideally, for a given starting time, it is preferable to include infinite (overtone) mode sum to completely characterize the total quadratic modes due to $(2,2,n)$ mode couplings. In practice, there are only finite linear modes that can be extracted from data with sufficiently accurate amplitude, i.e. a truncation has to be applied. For a ringdown analysis relevant for GW250114, we can faithfully predict the amplitudes of $(2,2,n)$ with $n\leq3$ at \starttime{} based on a weighted least-squares fitting towards numerical relativity or its surrogate waveform, such that a set of ten quadratic modes are available. We validate this truncation using the numerical-relativity simulation SXS:BBH:3617 \cite{Scheel:2025jct}, an equal-mass, non-spinning binary representative of GW250114. Smaller bundles of one-, three-, and six-\acp{QQNM} have not converged and change substantially as additional quadratic modes are included, similar to the behavior of truncated linear overtone sums \cite{Arnaudo:2025uos} (See \cref{fig:qqnm} in the Supplementary Materials). The leading omitted contribution involve the $(2,2,4)$ overtone. Its amplitude bounds the residual $(2,2,4)\times(2,2,n)$ couplings to a small fraction of the ten-modes summation at \starttime{}, with the mismatch between 10 modes and 15 modes description being \MismatchTenFifteenModes{}.
This is an indication that the first ten modes are already a faithful part of all QQNMs due to $(2,2,n)$ couplings at this time, without showing non-convergent signatures. 
}

\section{Results with GW250114}
\label{sec:results}

\reply{GW250114 is the loudest event detected so far. Given its nearly equal-mass nature, our analysis targets the \acp{QQNM} within the $(4,4)$ spherical-harmonic multipole augmented by the inspiral-merger physically informed prior. The inspiral and merger phases are modeled using \texttt{NRSur7dq4}, the most accurate waveform in the parameter region of GW250114 which includes all multipoles up to $\ell=4$, following the LVK choice for waveform \cite{LIGOScientific:2025obp}. The characteristic frequency and damping time of the parent linear \ac{QNM} $(2,2,n)$ are derived from the remnant mass and spin predicted by \texttt{NRSur7dq4}, while the corresponding amplitudes and phases are obtained via a weighted least-squares fit to the $(2,2)$ multipole. The \ac{QQNM} coupling coefficients are determined by second-order perturbation theory. The inference parameters in this analysis are precisely those that determine \texttt{NRSur7dq4}, including the binary masses, spins, sky location, distance, source orientation, polarization, and merger time and phase, introducing no additional degrees of freedom for the ringdown.
}

We analyze the interval $[-6,2]$ s of data \cite{LIGOScientific:2019lzm, KAGRA:2023pio, LIGOScientific:2025snk} centered on the merger time of GW250114  using \texttt{PyCBC Inference} \cite{Biwer_2019} in addition to the padding time at both ends of the data. To cleanly separate the inspiral-merger and ringdown stage, we apply gating and inpainting with durations of 2 and 1 s, removing inspiral and ringdown when analyzing the other region, respectively. The sky location and merger time are fixed to the maximum-likelihood values reported in \cite{KAGRA:2025oiz, LIGOScientific:2025obp}, with right ascension and declination $(\mathrm{ra, dec}) = (2.35,0.22)~\mathrm{rad}$ and the merger GPS time 1420878141.2362 s, \reply{where the merger time is defined as the peak of the quadrature sum of all available GW modes, as defined by \texttt{NRSur7dq4.}} The polarization angle is fixed to 1.37 rad. We use the \texttt{dynesty} \cite{speagle:2019} sampler with 30000 live points to sample the posterior distribution and stop when the change of logarithm of Bayesian evidence is less than 0.1. Priors are chosen as follows: the chirp mass and mass ratio have a prior uniform in the source-frame component masses, the spins follow an isotropic distribution with uniform amplitude up to 0.3, as guided by the inference results of small spins in \cite{KAGRA:2025oiz, LIGOScientific:2025obp}, the distance is assigned a prior uniform in the comoving volume, and the source orientation is isotropic. We explore a range of transition times between the inspiral-merger and ringdown segments, discretized from 3 to 9 $M_f$ relative to the merger time of GW250114, in steps of $1~M_\mathrm{f}$.

At times earlier than 6 $M_\mathrm{f}$ after the merger, additional non-modal signals can interfere with a clean \reply{weighted} least‑square fit for the linear $(2,2,n)$ modes. We therefore reconstruct these modes at start times from 6 to 10 $M_\mathrm{f}$ in steps of 1 $M_\mathrm{f}$, and extrapolate the resulting amplitudes backward to earlier times. In practice, we sample from a Gaussian distribution whose mean and variance are estimated from the multiple extrapolated values, thereby marginalizing over the uncertainties in the reconstruction. \reply{The stability of the parent mode amplitude fitting from multiple starting times is shown in \cref{app:consistency}}.

\begin{figure}[htbp]
\includegraphics[width=\columnwidth]{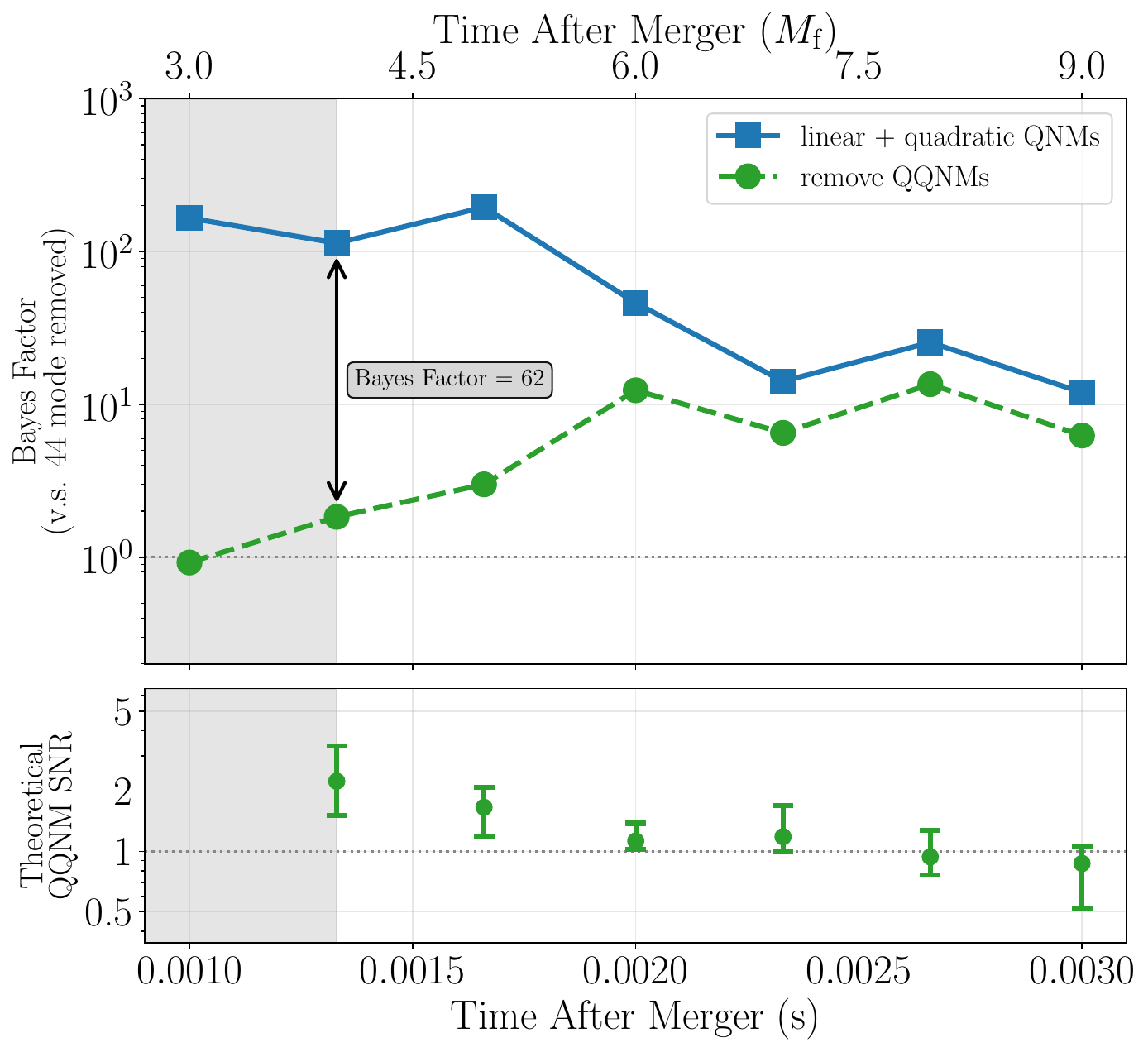}
\caption{
Top: Bayes factors of the fully nonlinear model and of the QQNM-subtracted model, each relative to the baseline with the entire $(4,4)$ multipole removed. The gap between the two curves quantifies the evidence for the quadratic modes, reaching \bayesf{} at \starttime{}. The shaded band marks start times earlier than \starttime{}, when the 10-mode QQNM bundle is not converged.
\reply{Bottom: Theoretical (optimal) network SNR of the ten quadratic mode bundle across the posterior samples. For start time below 6 $M_\mathrm{f}$, each sample's SNR is the median over draws of the Gaussian quadratic mode amplitude  estimated from the later 6-10 $M_\mathrm{f}$ reconstruction, as used in the parameter estimation. The SNR at $3 M_\mathrm{f}$ is not shown because of non-convergence of the QQNMs. At \starttime{} the median SNR is \expectedsnr{}.}
}
\label{fig:bf}
\end{figure}

To quantify the contribution of the quadratic modes, we construct two competing ringdown models. In the QQNM-subtracted model, the targeted nonlinear component is removed from the total ringdown signal, $h_{\rm RD} \rightarrow h_{\rm RD}-h_{\rm QQNM}$, where $h_{\rm QQNM}$ denotes the sum of the ten quadratic modes considered in this analysis. In the fully nonlinear model, the complete ringdown signal is retained. As a reference, we also define a baseline model in which the $(4,4)$ multipole is removed from \texttt{NRSur7dq4}. The Bayes factor between the fully nonlinear and QQNM-subtracted models then quantifies the evidence for the targeted quadratic modes.

\reply{In \cref{fig:bf}, we show the Bayes factors of the competing models relative to the baseline. At a ringdown start time of \starttime{} after the merger, the fully nonlinear model is favored over the QQNM-subtracted model with a Bayes factor of \bayesf{}. Across the explored range $3$--$9\,M_\mathrm{f}$, removing the quadratic modes consistently lowers the Bayesian evidence. The bottom panel shows the theoretical (optimal) network \ac{SNR} of the ten quadratic modes for all posterior samples. For starting time 5 $M_\mathrm{f}$ and earlier, we use the median of the Gaussian parent-amplitude estimation. At \starttime{}, the median is \expectedsnr{} and the 1 $\sigma$ posterior interval is \SNROneSigma{}. The curve is shown from 4 to 9 $M_\mathrm{f}$ because at 3 $M_\mathrm{f}$ the quadratic mode sum does not converge. The Bayes factor corresponds to an effective \ac{SNR} $\sqrt{2\ln\mathcal{B}} \simeq 2.9$, which lies within this interval.}

\begin{figure}[htbp]
\includegraphics[width=\columnwidth]{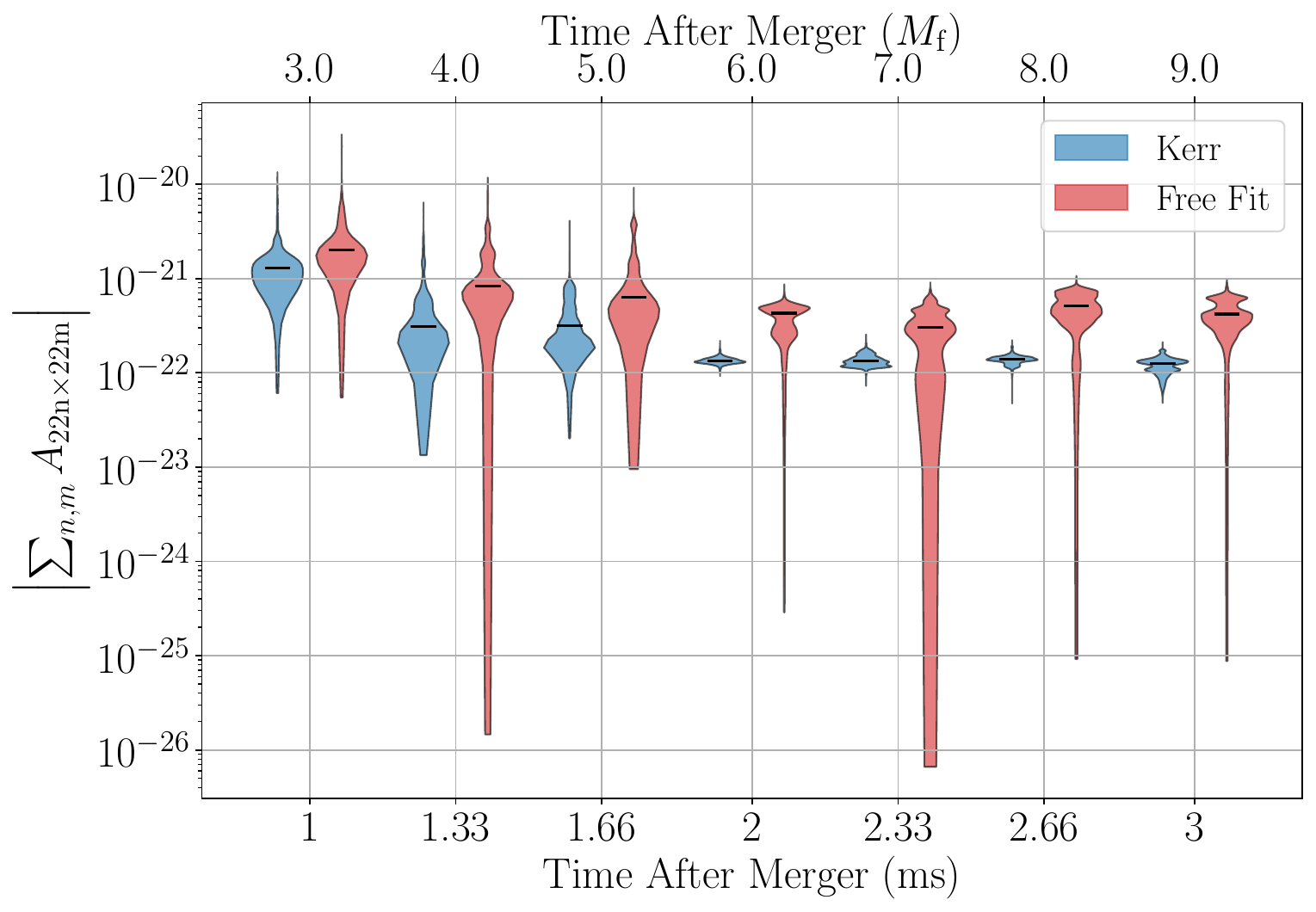}
\caption{
\reply{Total QQNM amplitude $|\sum_{n,m} A_{22n\times22m}|$ evaluated at each ringdown start time. The vertical axis is the instantaneous amplitude of the coherent sum at that time. Blue violins: \ac{GR} prediction from the posterior samples. Red violins: freely fitted amplitude $A_\mathrm{GR}(1-\delta A)$. Injection studies verify the free fit is consistent with the GR prediction.}
}
\label{fig:damplitude}
\end{figure}

\reply{The quadratic mode represents a crucial nonlinear \ac{GR} effect in the \ac{BH} ringdown. We test the consistency of its amplitude with the theoretical prediction by introducing a fractional deviation parameter $\delta A$, defined through $h_\mathrm{RD} \rightarrow h_\mathrm{RD} - \delta{A} \times h_\mathrm{QQNM}$. A value of $\delta A = 0$ corresponds to the \ac{GR} prediction and $\delta A = 1$ removes the quadratic modes entirely. Equivalently, the recovered amplitude for QQNM in units of the \ac{GR} prediction is $A/A_\mathrm{GR} = 1-\delta A$. As shown in \cref{fig:damplitude}, at \starttime{} the posterior gives $A/A_\mathrm{GR} = 3.1^{+0.9}_{-2.2}$ (the \ac{GR} value at the 7\% quantile) and at $5~M_{\rm f}$ it is $2.4^{+2.4}_{-1.6}$ (\ac{GR} at the 8\% quantile). Injection studies using the identical pipeline show no significant offset for a zero-noise \ac{GR} injection. In real detector noise, 24\% of the \ac{GR} injections produce a GR-value at an equally or more extreme posterior quantile as that observed for GW250114. The observed offset is therefore compatible with the real-noise injection distribution (see \cref{sec:app:injection}). An independent analysis \cite{Grimaldi:2026prn} with the parameterized effective-one-body framework finds the same mild preference for an amplitude above the \ac{GR} prediction as us. Importantly, a zero amplitude of the quadratic modes is excluded at \quasigma{}.}

\begin{figure}[htbp]
\includegraphics[width=\columnwidth]{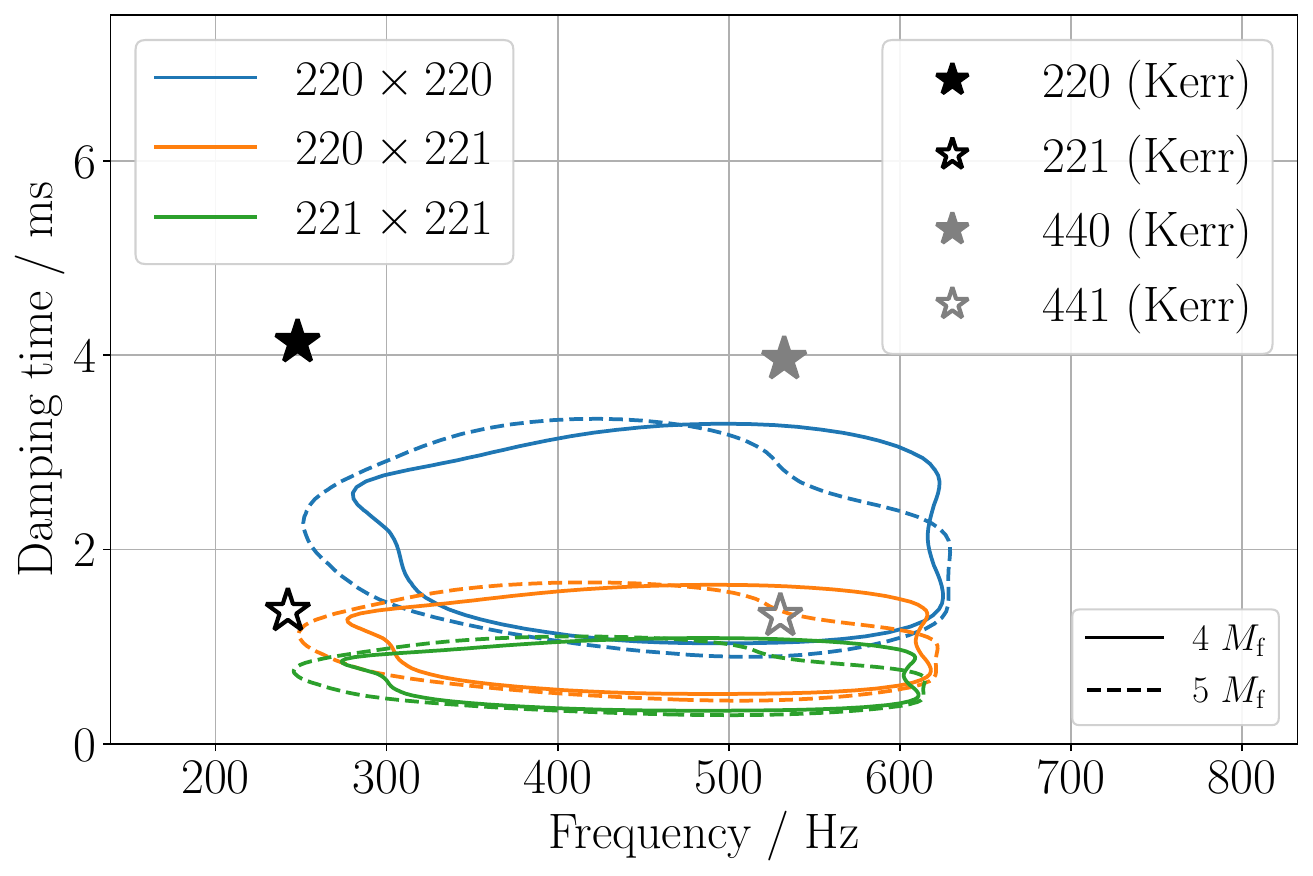}
\caption{\reply{The 90\% credible contours in the frequency--damping-time plane for three representative \acp{QQNM}. Colors identify the parent-mode pair: $220\times220$, $220\times221$, and $221\times221$. Solid and dashed contours show analyses starting at 4 and 5 $M_\mathrm{f}$, respectively. Filled and open stars mark the Kerr predictions for the fundamental and first overtone of the linear $(2,2)$ and $(4,4)$ families at the inferred remnant parameters. The quadratic-mode contours at both start times are consistent with the \ac{GR} predictions and disfavor the linear $(4,4,0)$, $(2,2,0)$, or $(2,2,1)$ interpretation (see text).}}
\label{fig:dfreqtau}
\end{figure}

\reply{We further measure the complex frequency of the \acp{QQNM} by allowing common fractional shifts, $f_\alpha = f_\alpha^\mathrm{GR}(1+\delta f)$ and $\tau_\alpha = \tau_\alpha^\mathrm{GR}(1+\delta \tau)$. At \starttime{}, the posterior gives $\delta f = -0.05^{+0.18}_{-0.22}$ and $\delta \tau = 0.01^{+0.39}_{-0.36}$; at $5\,M_{\rm f}$, it gives $\delta f = -0.10^{+0.33}_{-0.19}$ and $\delta \tau = -0.03^{+0.49}_{-0.33}$. The \ac{GR} value lies at the 47\% (\starttime{}) joint credible level and at the boundary of the 90\% contour (5 $M_\mathrm{f}$), with both marginal posteriors consistent with 0. \Cref{fig:dfreqtau} shows these contours in the frequency--damping-time plane, with 4 $M_\mathrm{f}$ shown by solid lines and 5 $M_\mathrm{f}$ by dashed lines, compared with the Kerr predictions for the linear $(2,2,0), (2,2,1), (4,4,0)$ and $(4,4,1)$ modes. The linear modes $(2,2,0),(2,2,1)$ and $(4,4,0)$ are disfavored, whereas $(4,4,1)$ is not immediately ruled out. However, we have performed an additional Bayes model comparison between the GR full waveform model and a model replacing the ten-mode QQNMs with a free-amplitude $(4,4,1)$ mode. The result shows that the QQNM is favored by a Bayes factor $ \mathcal{B}\simeq 11$ relative to a single $(4,4,1)$ mode at its Kerr frequency.}

\section{Discussion and Conclusions}\label{sec:con}
\label{sec:dis}

In this work, we report the statistical evidence of a set of ten \acp{QQNM} arising from the coupling among the fundamental $(2,2,0)$ mode and its first three overtones in the ringdown signal of GW250114. By incorporating information from the inspiral-merger phase into the ringdown analysis, we obtain a Bayes factor of \bayesf{} in favor of the quadratic mode model starting at \starttime{} after the merger. This analysis provides the first observational evidence for the excitation of quadratic modes in the remnant of a binary black hole merger. We establish a framework to separate quadratic modes from a superposition of multiple components close to the merger by theoretically computing their contributions. Inference of the amplitude \reply{and complex frequency} of the quadratic modes shows no deviation from theoretical expectations under a phenomenologically parameterized deviation, and the zero-quadratic-mode scenario is ruled out with significance of \quasigma.

Our work introduces a new ringdown analysis approach using the posterior from the inspiral-merger phase as a physically informed prior for the ringdown analysis.
This strategy enhances the detectability of subdominant ringdown modes by providing more constraining information from the pre-ringdown data.

Several future developments could further improve this framework.
These include marginalizing over the ringdown start time to better assess systematic uncertainties using methods such as those in \cite{Correia:2023bfn, Correia:2023ipz}, though such an approach requires a more complicated computation, thus beyond the scope of the current work.
In this work we assume the aligned-spin symmetry between the $+/-m$ mode in the quadratic modes, which is proper for GW250114 in light of its consistency with zero spin (see \cref{app:consistency}), but this can be extended to consider precessing spin in the future \cite{Zhu:2023fnf}.
Extending the analysis to earlier times will require modeling additional nonlinearities, such as the direct wave contribution \cite{Lu:2025vol,Oshita:2025qmn} and the dynamic excitation of modes \cite{Chavda:2024awq,DeAmicis:2025xuh}.
Applying the present method to additional events and combining evidence by stacking \cite{Yang:2017zxs} will also help detect fainter ringdown signals and improve the constraints on potential deviations from \ac{GR}.
The current parameterization for testing \ac{GR} employs a simple fractional deviation and assumes a \ac{GR} prior for the inspiral-merger phase, this framework can be refined in light of improved theoretical models.

The statistical evidence favors  the presence of \acp{QQNM} in GW250114, indicating a milestone that allows the identification of nonlinearities in a perturbed \reply{black} hole. Future observations are expected to detect a richer set of modal and non-modal, linear and nonlinear signatures in the ringdown. While this poses significant challenges for both theoretical studies and data analysis, it offers tremendous potential in understanding the fundamental predictions of \ac{GR} in the strong-field regime.

All data analysis results and scripts necessary to reproduce this work are released in \url{https://github.com/yi-fan-wang/nonlinear-ringdown-GW250114}.

\begin{acknowledgements}
The computational work for this manuscript was carried out on the Hypatia computer cluster at the Max Planck Institute for Gravitational Physics (Albert Einstein Institute) in Potsdam.
YFW thanks Collin Capano and Alex Correia for their development of the hierarchical model in \texttt{PyCBC}, which this work employs to integrate inspiral-merger inference with ringdown analysis, thanks Qian Hu for reviewing this manuscript in the LIGO Publications \& Presentations system, and thanks the Ringdown Inside and Out meeting organized by the Strong group at the Niels Bohr Institute, where the collaboration of this project was initiated.
This research has made use of data or software obtained from the Gravitational Wave Open Science Center (gwosc.org), a service of the LIGO Scientific Collaboration, the Virgo Collaboration, and KAGRA. This material is based upon work supported by NSF's LIGO Laboratory which is a major facility fully funded by the National Science Foundation, as well as the Science and Technology Facilities Council (STFC) of the United Kingdom, the Max-Planck-Society (MPS), and the State of Niedersachsen/Germany for support of the construction of Advanced LIGO and construction and operation of the GEO600 detector. Additional support for Advanced LIGO was provided by the Australian Research Council. Virgo is funded, through the European Gravitational Observatory (EGO), by the French Centre National de Recherche Scientifique (CNRS), the Italian Istituto Nazionale di Fisica Nucleare (INFN) and the Dutch Nikhef, with contributions by institutions from Belgium, Germany, Greece, Hungary, Ireland, Japan, Monaco, Poland, Portugal, Spain. KAGRA is supported by Ministry of Education, Culture, Sports, Science and Technology (MEXT), Japan Society for the Promotion of Science (JSPS) in Japan; National Research Foundation (NRF) and Ministry of Science and ICT (MSIT) in Korea; Academia Sinica (AS) and National Science and Technology Council (NSTC) in Taiwan.
Research at Perimeter Institute is supported in part by the Government of Canada through the Department of Innovation, Science and Economic Development and by the Province of Ontario through the Ministry of Colleges and Universities.
\end{acknowledgements}

\appendix

\reply{
\section{Statistical framework for QQNM detection and parameter estimation}
\label{sec:app:statistics}

\begin{figure*}[htbp]
\includegraphics[width=2\columnwidth]{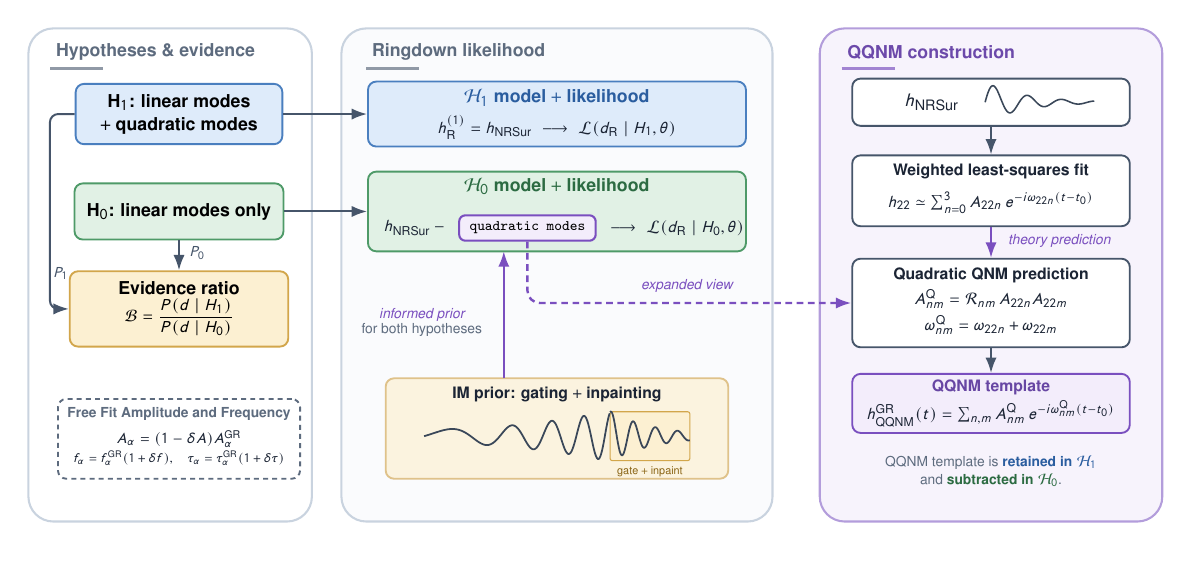}
\caption{A workflow for detecting \acp{QQNM}. The inspiral-merger-ringdown data are separated into inspiral-merger and ringdown segments, respectively, using gating and inpainting. The inspiral-merger posterior provides a physically informed prior for the ringdown likelihood.  Together with perturbative quadratic-coupling coefficients, these inputs define the predicted quadratic sector. Bayesian model selection between hypotheses with and without the target quadratic modes quantifies the evidence for their presence. A complementary free-fit consistency test framework is used to infer the quadratic-mode amplitude and frequency.
}
\label{fig:flowchart}
\end{figure*}

Ringdown analyses of current \ac{GW} events are entering a qualitatively new regime where several post-merger components can contribute comparably.  Near the merger, the signal can contain multiple overlapping contributions, including fundamental \acp{QNM}, overtones, and nonlinear quadratic modes, without a clean hierarchy in their contributions. In this situation, the traditional strategy of testing a single additional mode on top of a dominant background becomes insufficient. 

Here, we describe the statistical framework developed in this work, extending the standard Bayesian analysis for ringdown spectroscopy in two directions. First, we construct the signal detection framework in the presence of multiple ringdown components. Second, we incorporate inspiral-merger inference results as physically motivated priors for the ringdown phase. The same idea to integrate the information from inspiral and merger has been implemented by a parameterized effective-one-body waveform framework \cite{Brito:2018rfr,Ghosh:2021mrv,Maggio:2022hre,Pompili:2025cdc,Grimaldi:2026prn}. Combined with theoretical calculations of nonlinear mode couplings, this allows us to test for \acp{QQNM} without introducing independent free amplitude or phase parameters for our targets. We emphasize that this methodology is designed specifically for the early post-merger regime relevant for GW250114, where linear and nonlinear signals coexist and standard single-mode detection strategies become inadequate.

\subsection{Detection of multi-modes in ringdown}

We start by denoting the \ac{GW} data $d$ as the sum of the detector noise $n$ and \ac{GW} signals $h$, that is, $d = n + h$.
We first consider a simplified single-mode ringdown signal: $h(A, \omega) = A e^{-i \omega t}$, where $A$ is the complex amplitude at the starting time of ringdown, and $\omega$ is the complex eigen-frequency of a \ac{QNM}.
One way to quantify the detection of such a mode is to adopt the Bayesian model selection framework
\begin{align}
\label{eq:singlemode}
    \textit{Null Hypothesis } H_0 &: d = n, \nonumber \\
    \textit{Alternative Hypothesis } H_1 &: d = n + h(A,\omega) = n + A e^{-i \omega t}.
\end{align}
The odds ratio can be defined as
\begin{equation}    
\mathcal{O} := \frac{P(H_1 \mid d)}{P(H_0 \mid d)} = \frac{P(d \mid H_1)}{P(d \mid H_0)} \frac{P(H_1)}{P(H_0)} .
\end{equation}
Choosing an equal weight for the prior of $H_0$ and $H_1$, the odds ratio reduces to the Bayes factor 
\begin{equation}
\mathcal{B} = \frac{P(d \mid H_1)}{P(d \mid H_0)}.   \label{eq:app:bayesfactor}
\end{equation}
According to Jeffreys' scale, $\ln \mathcal{B} > 2.3$ ($\mathcal{B} > 10.0$) indicates strong evidence, $\ln \mathcal{B} > 3.5$ ($\mathcal{B} > 33.1$) very strong evidence, and $\ln \mathcal{B} > 4.6$ ($\mathcal{B} > 99.5$) is considered decisive evidence for the presence of a signal \cite{jeffreys1998theory}.
According to Bayes theorem, the posterior probability distribution of parameters $\vartheta$ given data $d$ is 
\begin{equation}
p(\vartheta \mid d, H) = \frac{\mathcal{L}(d \mid \vartheta, H) \pi(\vartheta \mid H)}{P(d \mid H)}
\end{equation}
where $\mathcal{L}(d|\vartheta)$ is the likelihood and $\pi(\vartheta)$ is the prior. The Bayes evidence can be computed by integrating the product of the likelihood and the prior. For example, if the prior distributions of $A$ and $\omega$ are $\pi(A), \pi(\omega)$, then 
\begin{equation}
\label{eq:app:compute_bayesfactor}
    P(d \mid H_1) =\int \mathrm{d}^2A \int \mathrm{d}^2\omega \mathcal{L}(d | A,\omega) \pi(A) \pi(\omega)
\end{equation}
where the integration over complex amplitude and complex frequency is performed.

As seen in \cref{eq:app:compute_bayesfactor}, prior knowledge does affect the value of the Bayes factor. 
However, the increase in significance occurs only when the ringdown data themselves are consistent with the inspiral-merger prediction (i.e., a larger likelihood), whereas inconsistent ringdown data would instead reduce the Bayesian evidence.
The prior distribution in the ringdown data analysis is usually determined by theoretical expectation, as shown in previous work \cite{LIGOScientific:2026wpt}. 
In our study of GW250114, the prior knowledge for the ringdown analysis is also informed by the data from the inspiral and merger (see below). 

With continuous improvement in signal \ac{SNR}, we are entering an era in which multiple interesting signals (overtones, quadratic modes, etc) are present and potentially detectable. GW250114 is likely the {\it first} event that allows the detection of a particular quadratic-mode signal while other ringdown signals are also significant.

We turn to consider multiple signals in the ringdown stage. Let us assume that 
\begin{align}
    d =n + \sum^N_{i=1} h_i\,,
\end{align}
with an integer $N >1$ and $h_i = A_i e^{-i \omega_i t}$ (this discussion can also be extended to non-modal signals).
A natural way to quantify the detection of a specific mode $h_N$, extending the single-mode method, is to perform Bayesian model selection for two hypotheses: 
\begin{align}
\label{eq:multiplemodes}
    \textit{Null Hypothesis } H_0 &: d =n+\sum^{N-1}_{i=1} h_i, \nonumber \\
    \textit{Alternative Hypothesis } H_1 &: d =n+\sum^{N}_{i=1} h_i.
\end{align}
If $N=1$, \cref{eq:multiplemodes} reduces to \cref{eq:singlemode}. 
If we identify the full theoretical waveform $\sum^{N}_{i=1} h_i$ as $h_{\rm thr}$, then the Bayes factor can be written as
\begin{equation}
    \mathcal{B} := \frac{P(d= n+h_{\rm thr} \mid H_1)}{P(d=n+h_{\rm thr}-h_N \mid H_0)}.
\end{equation}
Notice that this is exactly the same approach that we have adopted in the detection analysis for quadratic modes, with $h_N$ being the combined bundle of quadratic modes. 

In particular, a conventional approach would treat $h_N$ as an unconstrained additional waveform component with freely varying amplitudes and phases. Our framework differs fundamentally from this strategy. The nonlinear contribution is not introduced as an arbitrary extension of the waveform model. Instead, its amplitudes and phases are determined by:

\begin{enumerate}
\item the reconstructed parent linear modes via a weighted least-squares fit (see \cref{app:least-square}), and
\item theoretically computed quadratic coupling coefficients (see \cref{app:qqnm}).
\end{enumerate}

Consequently, the nonlinear \ac{QQNM} is not given independent amplitude and phase freedom in the ringdown inference. The test for \ac{QQNM} can instead be viewed as using the entire inspiral-merger-ringdown data, not ringdown alone. The Bayesian model selection therefore quantifies whether the inspiral-merger-ringdown data support the presence of specific \acp{QQNM} predicted by theory. A corresponding flowchart is shown in \cref{fig:flowchart}.
}

\subsection{Informing ringdown analysis with inspiral and merger}

\begin{figure}[htbp]
\includegraphics[width=\columnwidth]{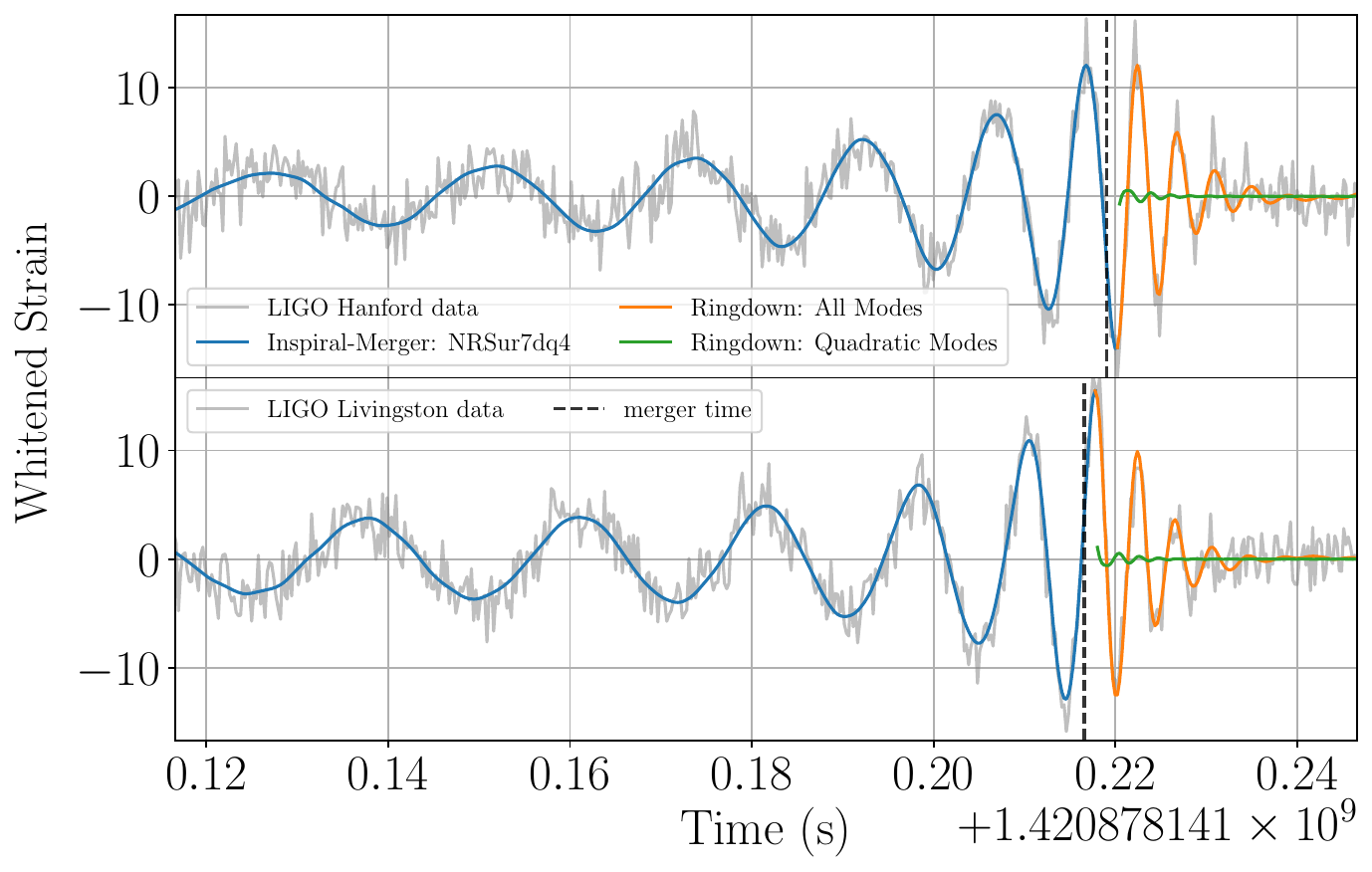}
\caption{Whitened strain data from LIGO Hanford (top) and Livingston (bottom) are shown in grey. 
The blue curves show the NRSur7dq4 inspiral--merger waveform, while the orange curves show the corresponding all mode ringdown beginning at \starttime{} after merger.
The green curves show the sum of ten \acp{QQNM} generated by couplings of the linear $(2,2,n)$ parent modes with $n\in\{0,1,2,3\}$.
All model waveforms are evaluated at the same maximum-likelihood source parameters inferred under the hypothesis excluding these ten predicted quadratic modes.
The vertical dashed line marks the merger time.
}
\label{fig:wf}
\end{figure}

A core component of our data analysis is integrating the inference results from the inspiral and merger stages to inform the ringdown analysis. We formalize this approach within a Bayesian framework for parameter estimation. We propose a composite likelihood following \cite{Correia:2023bfn,Correia:2023ipz} that cleanly separates the contributions from the inspiral-merger (IM) phase and the ringdown (R) phase
\begin{equation}
\mathcal{L}(\{d_\mathrm{IM}, d_\mathrm{R}\}|\vartheta) = \mathcal{L}_\mathrm{IM} (d_\mathrm{IM}|\vartheta)
\times
\mathcal{L}_\mathrm{R} (d_\mathrm{R}|\vartheta) 
\end{equation}
where $\vartheta$ denotes the model parameters. In this work, the parameters are the same in both stages, including the binary masses, spins, distance, sky location, orientation, polarization, coalescence phase and time that determine \texttt{NRSur7dq4}. The factorization is made possible by splitting the entire inspiral-merger-ringdown data series $d$ into two disjoint segments, $d_\mathrm{IM}$ and $d_\mathrm{R}$, using a gating-and-inpainting technique \cite{Capano:2021etf,Wang:2023ljx}.
\reply{The total Bayesian evidence can be written as:
\begin{equation}
P(\{d_\mathrm{IM}, d_\mathrm{R}\}) = P(d_\mathrm{IM}  ) \times P(d_\mathrm{R} \mid d_\mathrm{IM} ).
\end{equation}
where $P(d_\mathrm{R} \mid d_\mathrm{IM})$ is the Bayes evidence of the ringdown data conditional on the inspiral and merger data. The joint posterior can be reformulated as
\begin{equation}
p(\vartheta |\{d_\mathrm{IM}, d_\mathrm{R}\}) = 
\frac{\mathcal{L}_\mathrm{R} (d_\mathrm{R}|\vartheta)  
\times 
\Pi(\vartheta \mid d_\mathrm{IM})}
{P(d_\mathrm{R} \mid d_\mathrm{IM})}
\end{equation}
where we define the informed prior $\Pi(\vartheta \mid d_\mathrm{IM})$ as the posterior distribution informed from the inspiral-merger analysis:
\begin{equation}
    \Pi(\vartheta \mid d_\mathrm{IM})= \frac{\mathcal{L}_\mathrm{IM}(d_\mathrm{IM} \mid \vartheta) \times \pi(\vartheta)}{P(d_\mathrm{IM})}.
\end{equation}
}
In essence, we use gating and inpainting to isolate the likelihood contributions, then utilize the inspiral-merger inference results as a physically-motivated, informative prior for the subsequent ringdown-only analysis. Such a prior imposes strong constraints, to illustrate, imagine an extreme case in the limit of infinite \ac{SNR} during the inspiral-merger phase and arbitrarily accurate waveform modeling, we can predict the \ac{QNM} spectrum with arbitrary precision, thus providing an exact prior. As a result, this method enhances our detectability of sub-dominant \ac{QNM} modes, provided the ringdown data are consistent with the theoretical expectation. \reply{In practice, the constraining power of such a prior is limited by finite \ac{SNR}, hence statistical uncertainty, and potential modeling errors. We emphasize that this prior is derived from inference with the inspiral and merger data, reflecting the physics that ringdown excitation is determined by the initial condition set up by the inspiral and merger, without much room from human tuning.} An illustrative figure separating the inspiral-merger and ringdown whitened waveform is shown in \cref{fig:wf}.

\reply{
\subsection{A parameterization for free-fit amplitude and frequency}
\label{app:freefits}

In addition to the detection analysis, in which the amplitudes, phases, and frequencies of the quadratic modes are fully fixed by a parameterization of an inspiral-merger-ringdown waveform and by perturbation theory, it is instructive to relax these assumptions and infer these parameters with a more agnostic parameterization. To this end, we free the amplitude or the complex frequency of the quadratic sector through explicit deviation parameters and infer their posterior distributions. This checks whether the inference favored by the data is indeed consistent with the quadratic modes predicted by theory.

Returning to the single-mode example, the amplitude and frequency can be decomposed into the sum of the \ac{GR} prediction and deviations: $A=A_{\rm GR}+\delta A, \omega=\omega_{\rm GR}+\delta \omega$. 
In such a parameterization, the posterior distribution of $\delta A$ and $\delta \omega$ can be written as
\begin{align}
    P(\delta A,\delta \omega \mid d) =& \int d^2 A_{\rm GR} d^2\omega_{\rm GR}P(d \mid \{A,\omega\}) \nonumber \\ 
    &\times \pi (A_{\rm GR}, \omega_{\rm GR}, \delta_A, \delta \omega) P(d)^{-1}\,.
\end{align}
If there are $N$ signals present, we introduce the deviation from \ac{GR} for each of the modes $A_i=A_{i, \rm GR}+\delta A_i$, $\omega_i=\omega_{i, \rm GR}+\delta \omega_i$,
the joint posterior is
\begin{align}
P(\{\delta A_i,\delta \omega_i \} \mid d) =& \int  \left[ \prod^N_{i=1}d^2 A_{\rm i, GR} d^2\omega_{\rm i,GR} \right]  P(d \mid \{A_i,\omega_i\})  \nonumber \\
&\times \pi(A_{\rm i, GR}, \omega_{\rm i, GR}, \delta A_i, \delta \omega_i) P(d)^{-1}\,.
\end{align}
Such implementation relies on the identification of each signal $h_i$ and the accurate description in its \ac{GR} sector. The advantage is that it allows simultaneous parameterized deviation for all signals, such that the joint posterior provides a comprehensive description of the entire signal spectrum. This is the conventional method used to claim detection of a single ringdown mode. 

In practice, we have a relatively accurate waveform model for $h_{\rm thr}$ through a surrogate model of numerical simulations. Adopting independent deviation parameters $\delta A_i, \delta \omega_i$ is, however, not feasible considering a set of ten \acp{QQNM} in GW250114. The data can not constrain such a high-dimensional deviation parameter space, and the resulting posterior would not be constraining.

We therefore reduce the deviation space by adopting a common fractional deviation of all modes in the quadratic bundle, motivated by that we treat the bundle as a whole component. For the free fits of amplitude, we construct such a parameterization 
\begin{equation}
A_\alpha = (1-\delta A)\,A_{\alpha}^{\rm GR}
\end{equation}
where $\alpha$ denotes the index of ten \ac{QQNM}. The amplitude test adopts this component in the ringdown waveform, hence
\begin{equation}
    h_{\rm RD} \rightarrow
    h_{\rm RD} - \delta A \, h_{\rm QQNM}^{\rm GR},
\end{equation}
where $\delta A = 0$ recovers the \ac{GR} prediction and $\delta A = 1$ removes the quadratic modes entirely. The prior for $\delta A$ is chosen to be uniform in $[-5,1]$.

The spectral test shifts the frequencies and damping times through
\begin{equation}
    f_{\alpha} = f_{\alpha}^{\rm GR}\,(1+\delta f), \quad
    \tau_{\alpha} = \tau_{\alpha}^{\rm GR}\,(1+\delta \tau).
\end{equation}
The ringdown waveform is then modified by subtracting the \ac{GR} predicted \ac{QQNM} and adding back the \ac{QQNM} with free frequencies and damping times,
\begin{equation}
    h_{\rm RD} \rightarrow
    h_{\rm RD} -  h_{\rm QQNM}^{\rm GR} + h_{\rm QQNM}(\delta f, \delta \tau),
\end{equation} 
such that $\delta f = \delta \tau = 0 $ recovers the \ac{GR} prediction. The prior for $\delta f$ and $\delta \tau$ is chosen to be uniform in $[-0.5, 0.5]$.
In light of the observation that the Kerr frequency of $(4,4,1)$ is not immediately ruled out by the free frequency fit of QQNMs in \cref{fig:dfreqtau}, we also test the $(4,4,1)$ hypothesis by introducing the following model
\begin{equation}
       h_{\rm RD} \rightarrow
    h_{\rm RD} -  h_{\rm QQNM}^{\rm GR} + h_{441},
\end{equation}
where the $h_{441}$ damped sinusoid has the Kerr frequency of the linear $(4,4,1)$ mode and free amplitude and phase. The prior for the dimensionless amplitude $rA_{441}^{\rm free}/M_\mathrm{f}$ is uniform in $[0, 0.3]$, where $r$ is the luminosity distance and $M_\mathrm{f}$ is the remnant mass ($G=c=1$), the prior for phase is uniform in $[0,2\pi]$.  

\subsection{Relation between signal-to-noise ratio and Bayes factor}

The detectability of the quadratic bundle can be understood from the relation between the Bayesian evidence and the \ac{SNR}.
The optimal (theoretical) \ac{SNR} of the quadratic template $h_\mathrm{QQNM}$ is
\begin{equation}
    \rho_\mathrm{opt}^2 = (h_\mathrm{QQNM}|h_\mathrm{QQNM})
    = 4 \int_0^\infty
      \frac{|\tilde{h}_\mathrm{QQNM}(f)|^2}{S_n(f)}\,\mathrm{d}f,
\end{equation}
and its matched-filter \ac{SNR} with data is
\begin{equation}
    \rho_\mathrm{mf} =
    \frac{(d - h_\mathrm{linear}\,|\,h_\mathrm{QQNM})}
{\sqrt{(h_\mathrm{QQNM}|h_\mathrm{QQNM})}},
\end{equation}
where $S_n(f)$ is the one-sided noise power spectral density, $(\cdot|\cdot)$ is the noise-weighted inner product and
$h_\mathrm{thr} = h_\mathrm{linear} + h_\mathrm{QQNM}$ is the full waveform model.
For a fully specified template with no free parameters, the
log-likelihood ratio between the hypotheses with and without
$h_\mathrm{QQNM}$ is
\begin{equation}
    \ln \Lambda
    = \rho_\mathrm{mf}\,\rho_\mathrm{opt}
    - \tfrac{1}{2}\,\rho_\mathrm{opt}^2 ,
\end{equation}
and when the data contain the predicted signal, $\langle \rho_\mathrm{mf} \rangle = \rho_\mathrm{opt}$, so
$\langle \ln \Lambda \rangle = \rho_\mathrm{opt}^2/2$. For GW250114 at \starttime{}, the measured $\ln\mathcal{B}=\ln 62\simeq4.13$ corresponds to an effective \ac{SNR} $\sqrt{2\ln\mathcal{B}}\simeq2.87$. This value lies within the 1 $\sigma$ posterior interval of the nominal theoretical \ac{SNR} shown in \cref{fig:bf}.
}

\section{Computation of QQNM coupling coefficients}
\label{app:qqnm}
In this section we present a brief overview of the computation of the \ac{QQNM} coupling coefficients using the techniques described in Ref.~\cite{Khera:2024bjs}. 

Wave-wave interactions produce nonlinearities in the black hole ringdowns. While the tail signal is generically small and the direct wave primarily matters in the merger stage before 4 $M_f$ \cite{Oshita:2025qmn}, the interaction of \acp{QNM} should dominate after that point. At second-order, the coupling of $(2,2,n)$ modes to quadratic modes within the $(4,4)$ multipole \cite{Khera:2024bjs,Ma:2024qcv} is
\begin{align}\label{eq:qsum}
   & \sum_{n,m} rA_{22n\times22m} e^{-i \omega_{22n\times22m} t} \nonumber \\
    &= \sum_{n,m} \mathcal{R}_{22n\times22m}(a_f) \times rA_{22n} \times rA_{22m} e^{-i \omega_{22n} t} e^{-i \omega_{22m} t}
\end{align}
where $rA$ is the product of the source distance and the amplitude which is measured in unit of $M_f$. The subscripts $22n\times22m$ refer to the \ac{QQNM} from coupling between the two parent linear \acp{QNM} $(2,2,n)$ and $(2,2,m)$. The coupling coefficient $\mathcal{R}_{22n\times22m}$ depends on the remnant spin $a_f$ and has unit of $1/M_f$. 

Given a pair of QNMs with frequencies $\omega_1$ and $\omega_2$ with azimuthal numbers $m_1$ and $m_2$, at second order in perturbation theory these QNMs produce quadratic modes with frequencies $\omega_1 + \omega_2$, $\omega_1 - \bar{\omega}_2$, and $-\bar{\omega}_1 + \omega_2$ and azimuthal numbers $m_1+m_2$, $m_1-m_2$ and $-m_1+m_2$ respectively, where the bar denotes complex conjugate. The excitation of the combination $-\bar{\omega}_1 - \bar{\omega}_2$ happens to vanish, as shown in ~\cite{Khera:2024bjs}.

Note that in addition for every QNM with frequency $\omega_1$, there is another QNM with frequency $-\bar{\omega}_1$, which we call its conjugate mode. For example, the prograde $(2,2,0)$ and prograde $(2,-2,0)$ QNMs are conjugate modes whose frequencies are related as above. Consequently, any given quadratic frequency component can be generated by up to 3 combinations of QNM pairs. For instance, $\omega_{220} + \omega_{221}$ frequency component is generated by the products of the prograde modes $(2,2,0)\times(2,2,1)$, $(2,2,0)\times(2,-2,1)$ and $(2,-2,0)\times(2,2,1)$. The amplitude of this frequency component will be given by
\begin{align}
    A_{220\times221} &= \mathcal{R}_{++} A_{220}A_{221} \\ \nonumber 
    &+ \mathcal{R}_{+-}A_{220}\bar{A}_{2-21} + \mathcal{R}_{-+}\bar{A}_{2-20}A_{221}
\end{align}
However, if the system has reflection symmetry about the z-axis, such as for non-precessing systems, this relation simplifies significantly. The amplitude $A_{\ell m n}$ of the prograde $(\ell, m,n)$ mode is related to the prograde amplitude $A_{\ell(-m) n}$ of the prograde $(\ell, -m, n)$ mode by $A_{\ell m n} = (-1)^\ell \bar{A}_{\ell(-m) n}$. This allows us to combine all three contributions into a single amplitude ratio $\mathcal{R}$, which is used in this paper. For quadratic mode $(\ell_1,\,m_1,\,n_1) \times (\ell_2,\,m_2,\,n_2)$, the non-precessing quadratic ratio is given  by:
\begin{equation}
    \mathcal{R}= \mathcal{R}_{++}  + (-1)^{\ell_2}\mathcal{R}_{+-} + (-1)^{\ell_1} \mathcal{R}_{-+} \,.
\end{equation}

\begin{figure}[htbp]
\includegraphics[width=\columnwidth]{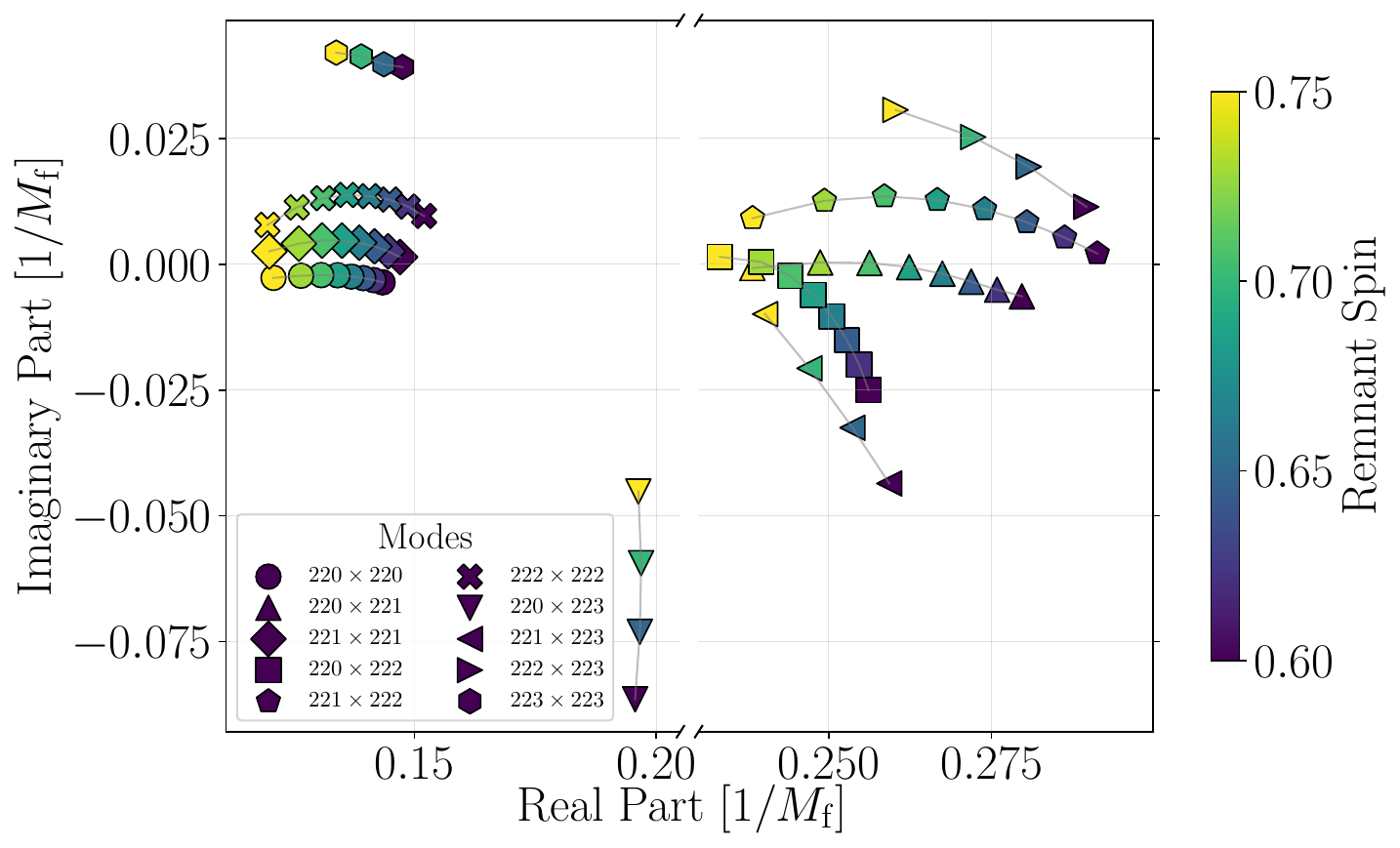}
\caption{The coupling coefficient $\mathcal{R}_{22n\times22m}$ from  different coupling channels as a function \reply{of the remnant spin}, obtained by solving the reduced second-order Teukolsky equation through a hyperboloidal slicing \cite{Khera:2024bjs}.
}
\label{fig:theory}
\end{figure}

The individual ratios $\mathcal{R}_{++}$, $\mathcal{R}_{+-}$ and $\mathcal{R}_{-+}$ are computed using pseudospectral methods on a hyperboloidal slicing. The steps to obtain the ratio are as follows: (i) Solve the homogeneous Teukolsky equation to obtain the linearized $\Psi_4^{(1)}$ for a pair of QNMs. This is done using the techniques described in Ref.~\cite{Ripley:2022ypi}; (ii) Perform metric reconstruction from $\Psi_4$ to obtain the corresponding linearized metric; (iii) Compute the source term for the reduced second order Teukolsky equation, a quadratic function of the linearized metric. The source term can be split into the $++$, $+-$, and $-+$ channels based on their frequencies, as described above; (iv) For each channel, solve the inhomogeneous Teukolsky equation for $\Psi_4$ with the source term and outgoing boundary conditions to obtain the second order $\Psi_4^{(2)}$ and perform metric reconstruction for the QNMs, compute the second order source term and solve the reduced second order Teukolsky equation. Lastly, at infinity, the reduced second order $\Psi_4^{(2)}$ is related to the second order gravitational radiation $h^{(2)}$ by $\Psi_4^{(2)} = -\ddot{h}/2$, allowing us to compute the amplitude of the \ac{QQNM}, and hence the amplitude ratios $R_{++}$, $R_{+-}$ and $R_{-+}$. For non-precessing binaries, the results are illustrated in \cref{fig:theory}.
The coupling coefficient is a function of the remnant spin only, and we interpolate precomputed tables over spin for each of the ten mode pairs.

\section{Computation of linear (2,2,n) modes' amplitudes}
\label{app:least-square}

\subsection{Weighted Least Square Fitting}
\reply{The amplitudes of the \acp{QQNM} targeted in this work are not free parameters and fixed by the parent linear $(2,2,n)$ modes' amplitudes and the theoretically computed quadratic coupling coefficients.}
The \reply{weighted} least-squares fitting is used to obtain the amplitude of the $(2,2,n)$ \ac{QNM} overtones from an inspiral-merger-ringdown numerical simulation or its surrogate model.
Consider a model composed of multiple \acp{QNM}:
\begin{equation}
    h(t) = \sum_{k=1}^{M} A_k e^{-i\omega_k t},
    \label{eq:multi_freq_model}
\end{equation}
where $A_k$ is the complex amplitude of the $k$-th mode, $\omega_k$ is the known complex frequency, and $M$ is the total number of modes.

The data to be decomposed, such as the $(2,2)$ multipole of a numerical relativity simulation or its surrogate, can be modeled as
\begin{equation}
    d(t) = h(t) + \Delta h(t) + n(t),
    \label{eq:data-budget}
\end{equation}
where $n(t)$ is the numerical noise of the simulation or surrogate, with
variance $\langle |n(t)|^2 \rangle = \epsilon^2(t)$, and $\Delta h(t)$ is the model error, any ringdown content absent from \cref{eq:multi_freq_model}, such as overtones beyond the truncation, whose envelope we denote $\langle |\Delta h(t)|^2 \rangle = \sigma^2(t)$.
We define the residual weighted by the inverse of the total variance,
\begin{equation}
    R = \int \frac{
    \left| d(t) - \sum_{k=1}^{M} A_k e^{-i\omega_k t} \right|^2
    }{\epsilon^2(t) + \sigma^2(t)} {\rm d}t,
    \label{eq:weighted-residual}
\end{equation}
The weighting factor accounts for the systematic error in the ringdown model and the numerical noise in the data.

The objective is to obtain a fit of the complex amplitudes $A_k$ to minimize $R$.
Define the data vector $\mathbf{d}$, amplitude vector $\mathbf{A}$, and matrix $\mathbf{G}$, $\mathbf{W}$

\begin{align}
    \mathbf{d} &= \begin{bmatrix} d(t_1) & d(t_2) & \cdots & d(t_N) \end{bmatrix}^T, \\
    \mathbf{A} &= \begin{bmatrix} A_1 & A_2 & \cdots & A_M \end{bmatrix}^T, \\
    \mathbf{G} &=
    \begin{bmatrix}
        e^{-i\omega_1 t_1} & e^{-i\omega_2 t_1} & \cdots & e^{-i\omega_M t_1} \\
        e^{-i\omega_1 t_2} & e^{-i\omega_2 t_2} & \cdots & e^{-i\omega_M t_2} \\
        \vdots & \vdots & \ddots & \vdots \\
        e^{-i\omega_1 t_N} & e^{-i\omega_2 t_N} & \cdots & e^{-i\omega_M t_N}
    \end{bmatrix}, \\
    \mathbf{W} &= \mathrm{diag}\!\left(
    \frac{1}{\epsilon^2(t_1)+\sigma^2(t_1)}, \cdots,
    \frac{1}{\epsilon^2(t_N)+\sigma^2(t_N)}\right),\end{align}
where the superscript $T$ denotes transpose. The \ac{QNM} waveform can be compactly written as
\begin{equation}
    \mathbf{h} = \mathbf{G} \mathbf{A}.
    \label{eq:matrix_model}
\end{equation}
The sum of squared residuals can be expressed as
\begin{equation}
    R = (\mathbf{d} - \mathbf{G}\mathbf{A})^H \mathbf{W} (\mathbf{d} - \mathbf{G}\mathbf{A}),
    \label{eq:matrix_residual}
\end{equation}
where the superscript $H$ denotes the conjugate transpose.
Taking the derivative with respect to the complex conjugate $\mathbf{A}^H$ and setting it to zero
\begin{equation}
    \frac{\partial R}{\partial \mathbf{A}^H} = -\mathbf{G}^H \mathbf{W} \mathbf{d} + \mathbf{G}^H \mathbf{W} \mathbf{G} \mathbf{A} = 0,
    \label{eq:normal_eq_deriv}
\end{equation}
we obtain the solution
\begin{equation}
    \mathbf{A} = (\mathbf{G}^H \mathbf{W} \mathbf{G})^{-1} \mathbf{G}^H \mathbf{W} \mathbf{d}.
    \label{eq:lsq_solution}
\end{equation}

The weights in \cref{eq:weighted-residual} require an estimate of $\sigma(t)$, which is dominated by the first overtone omitted from the fit. In this work the fitted modes are $(2,2,n)$ with $n \leq 3$, matching the parent modes entering the ten \acp{QQNM} of the truncated sum in \cref{eq:qsum}, and the leading omitted overtone is $(2,2,4)$. We perform a two-step procedure to determine the systematic error. First, an ordinary least-squares fit including $(2,2,4)$ provides an estimate of its amplitude $|A_{224}|$. The model-error envelope is then
\begin{equation}
    \sigma(t) = |A_{224}|\, e^{-(t - t_0)/\tau_{224}},
    \label{eq:sigma-envelope}
\end{equation}
with $\tau_{224}$ the Kerr damping time of the $(2,2,4)$ mode. Second, the modes $(2,2,n)$, $n \leq 3$, are refitted with the weights of \cref{eq:weighted-residual}, the $(2,2,4)$ mode is only used to determine the weighting factor and is not part of the final fit model. The noise floor $\epsilon$ is set to $\sim 1\%$ of the $(2,2)$ amplitude at the start of the fit window. We have verified that the fitting results are insensitive to this choice within roughly an order of magnitude.

\subsection{Consistency Test for Amplitude Extraction of the $(2,2,n)$ modes}
\label{app:consistency}

To validate the consistency of the extraction of \ac{QNM} amplitudes, \reply{we test whether amplitudes fitted at different start times are consistent at a common reference time}. We consider a $(30,30)~M_\odot$ binary with zero spin which is representative of GW250114, and extract the $(2,2,n)$ \acp{QNM} ($n\leq3$) from \texttt{NRSur7dq4} at multiple starting times from 6 to 10~$M_\mathrm{f}$ after the merger. The \reply{extracted} amplitudes are then extrapolated back to \starttime{} \reply{using the damped-sinusoid model with Kerr frequencies and decay times determined from the remnant mass and spin}, the reference time adopted in this work to report the detection of a set of ten \acp{QQNM}. For a perfect model the amplitude fits from multiple starting times should coincide. Their scatter, quantified by the ratio of the standard deviation to the mean, provides a measure of the fitting fidelity.

\begin{figure}[htbp]
\includegraphics[width=\columnwidth]{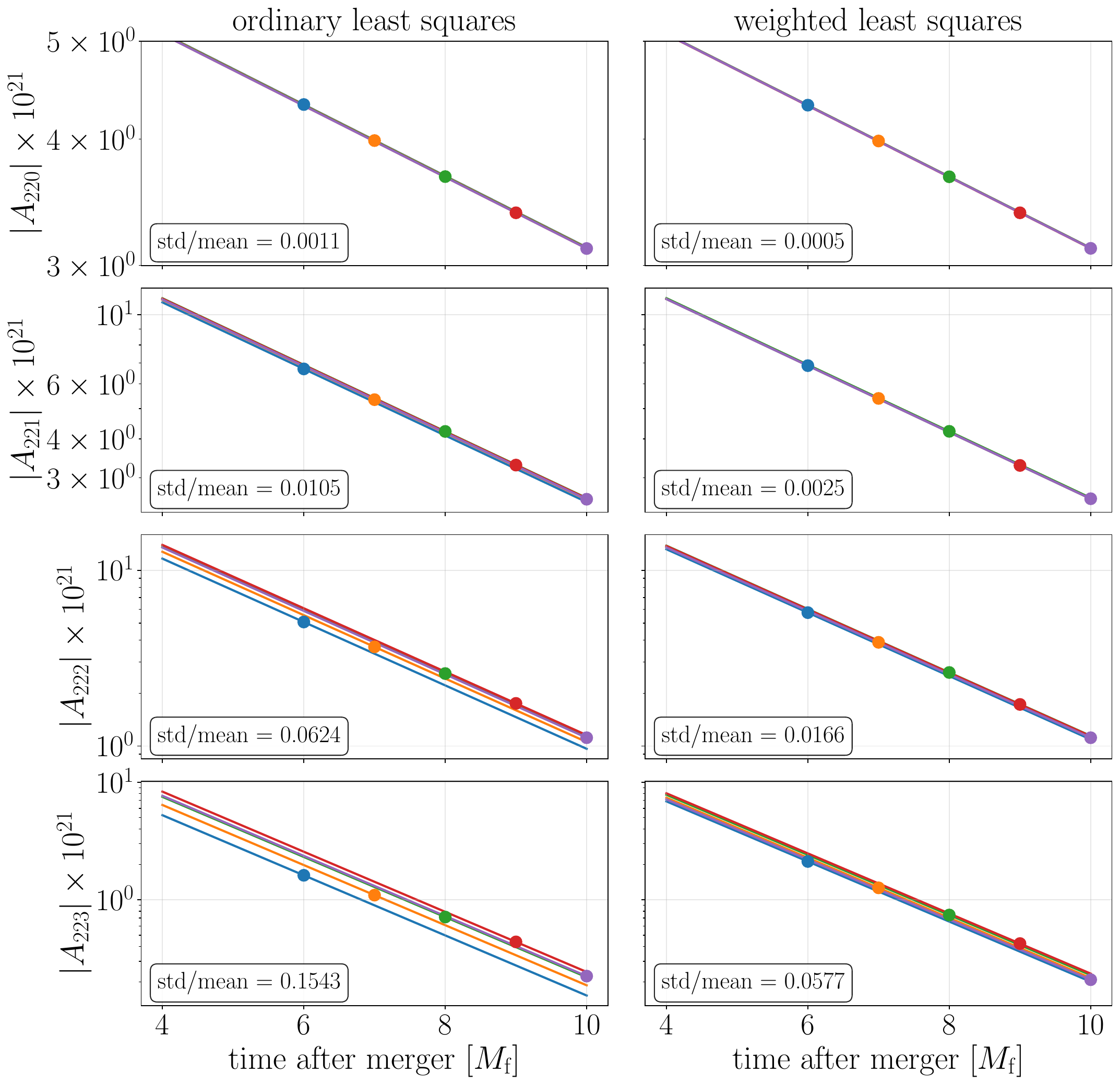}
\caption{\reply{Comparison of the $(2,2,n)$ amplitudes extracted from fitting windows that begin at different post-merger times}, for ordinary (left) and weighted (right) least squares. Dots show the amplitudes obtained from fitting windows beginning at $6$, $7$, $8$, $9$, and $10~M_\mathrm{f}$ after the merger (one color per fitting-window choice), and solid lines their extrapolation with the Kerr damped-sinusoid model. The bundle of extrapolated curves is visibly tighter for the weighted fit, the ratio of the standard deviation to the mean of the amplitudes extrapolated to \reply{$4~M_\mathrm{f}$} (left end of each curve) is annotated in each panel.}
\label{fig:weighted-consistency}
\end{figure}

\Cref{fig:weighted-consistency} compares the weighted least-squares fit with the ordinary least-squares fit. The weighted fit improves the \reply{consistency among fitting windows beginning at different post-merger times} of every mode by factors of $2.3$, $4.2$, $3.8$, and $2.7$ for $(2,2,0)$ through $(2,2,3)$, respectively. The corresponding \reply{scatter among the five fits whose windows begin at $6$, $7$, $8$, $9$, and $10\,M_\mathrm{f}$} is $0.046\%$, $0.252\%$, $1.657\%$, and $5.771\%$. These results demonstrate that the extrapolated amplitudes at \starttime{} are robust against \reply{changing the post-merger time at which the fitting window begins}. 

\reply{We further test the systematics by assessing possible $(3,2,0)$ contamination using the equal-mass, nonspinning $(30,30)\,M_\odot$ \texttt{NRSur7dq4} waveform. At $4\,M_\mathrm{f}$, adding the $(3,2,0)$ mode to the weighted least-squares fit to the $(2,2)$ multipole gives $|A_{320}|/|A_{220}|=3.8\times10^{-3}$. This changes the ten-mode quadratic sum by a mismatch of $5.1\times10^{-3}$. We therefore treat this component as part of the surrogate-modeling noise floor.}

\reply{Finally, our construction of the $(4,\pm4)$ waveform sector assumes the nonprecessing symmetry $h_{\ell,-m}=(-1)^\ell h_{\ell m}^{*}$. We assess this approximation using an \texttt{NRSur7dq4} waveform generated from the maximum-likelihood sample of the public LVK GW250114 posterior, which has $\chi_p=0.134$. Replacing the true precessing $h_{4,-4}$ mode with the symmetric approximation $h_{44}^{*}$ produces a mismatch of $1\times10^{-5}$. We also compare this waveform with an aligned-spin waveform obtained by setting the in-plane spin components to zero. At $4\,M_\mathrm{f}$, the fitted $(2,2,n)$ amplitudes differ by $0.5\%$, $3.6\%$, $5.5\%$, and $8.2\%$ for $n=0-3$, the ten-mode QQNM sum changes with a mismatch of $0.0376$. These systematics remain below the approximately $ 1/\expectedsnr = 45\%$ statistical amplitude uncertainty associated with the theoretical \ac{SNR} $\sim\expectedsnr$ and are therefore not expected to affect the model-selection conclusion.}

\section{Anatomy of the (4,4) multipole}
\label{sec:app:anatomy}

In the main text we have used the sum of ten quadratic modes as a bundle and \starttime{} as a starting time for the ringdown analysis. The primary motivation is ten-mode \ac{QQNM} comprises a dominant part of the quadratic modes at the analyzed starting time. In this section, we further present the supporting studies on the modal/non-modal content of the $(4,4)$ multipole signal, in particular the linear modes, i.e. $(4,4,0), (4,4,1)$. We emphasize that the Bayesian model selection does not require the detailed characterization of the residual signal, nevertheless it is still instructive to diagnose the signal contents. All analyses use a numerical relativity simulation SXS:BBH:3617 \cite{Scheel:2025jct}, an equal-mass, non-spinning binary black hole merger representative of GW250114. 

\subsection{Convergence of the truncated \ac{QQNM} sum}

The \ac{QQNM} within the $(4,4)$ multipole is formally an infinite sum over parent overtones in \cref{eq:qsum}. In practice, the sum must be truncated at the highest parent mode whose amplitude can still be reconstructed reliably. The weighted least-squares fit in \cref{app:least-square} is accurate for parents up to $(2,2,3)$, which defines the ten-mode bundle. The leading omitted couplings involve the $(2,2,4)$ overtone. Fitting $(2,2,4)$ directly from the numerical simulation bounds the residual $(2,2,4)\times(2,2,n)$ couplings to a small fraction of the ten-mode sum at \starttime{}.

\begin{figure}[htbp]
\includegraphics[width=\columnwidth]{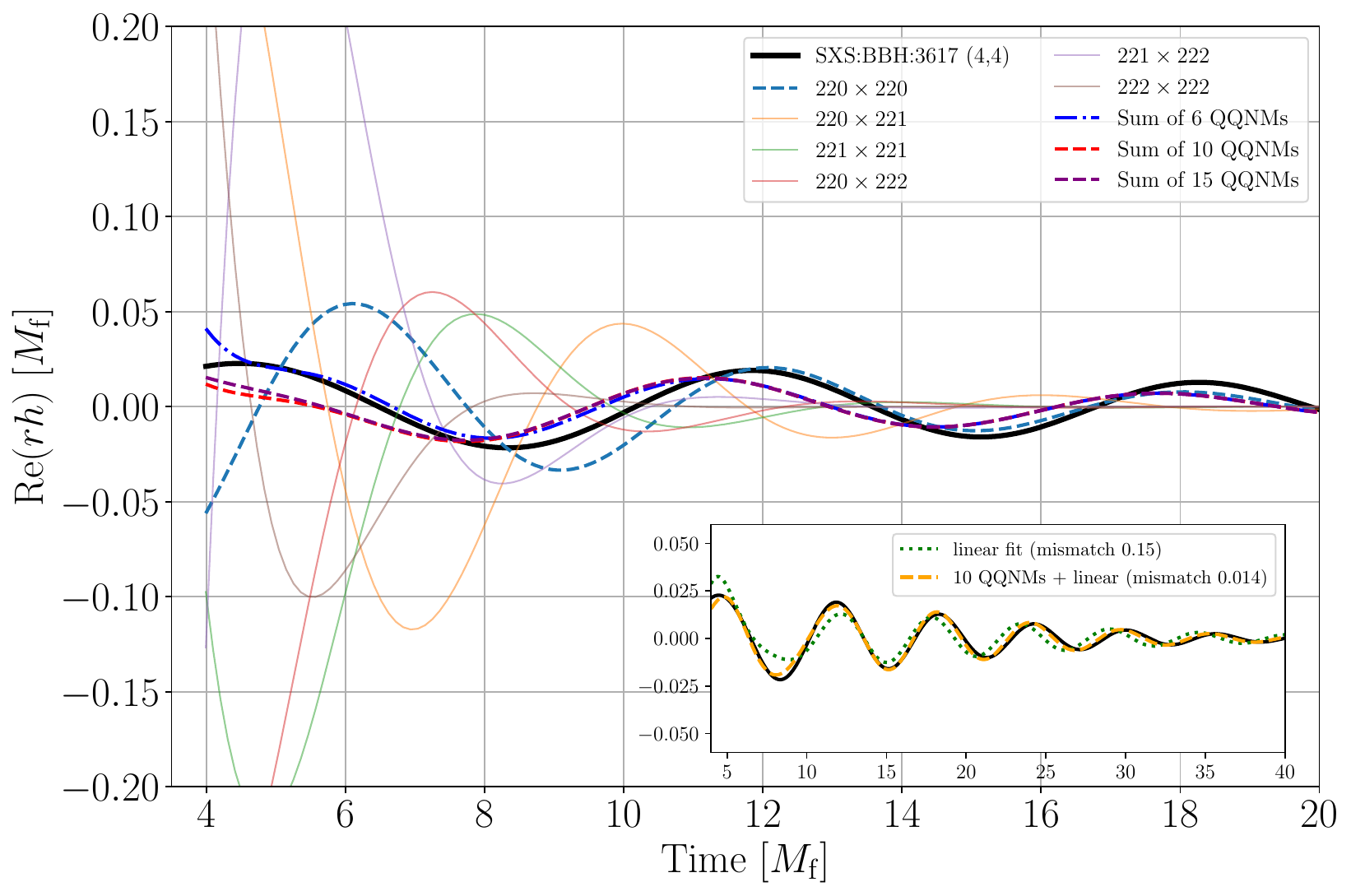}
\caption{Comparison of the $(4,4)$ multipole of the numerical relativity simulations SXS:BBH:3617 and various quadratic modes.
\reply{The six-, ten-, and fifteen-mode sums include parent overtones through $(2,2,2)$, $(2,2,3)$, and $(2,2,4)$, respectively. The agreement between the ten- and fifteen-mode waveforms at $4\,M_\mathrm{f}$ shows convergence of the truncation. The inset compares a linear $(4,4,0)+(4,4,1)$ fit with the reconstruction obtained after adding the ten predicted quadratic modes.}
}
\label{fig:qqnm}
\end{figure}

\Cref{fig:qqnm} compares the $(4,4)$ multipole with partial sums of the predicted \acp{QQNM}. The partial sums vary strongly as modes are added: the one-, three-, and six-mode truncations are divergent from the total signal, and settle only at ten modes. This behavior is similar to the non-convergence of truncated linear overtone sums \cite{Arnaudo:2025uos}. 
As shown in the inset of \cref{fig:qqnm}, fitting only the linear $(4,4,0)$ and $(4,4,1)$ modes to the full multipole reaches a match of $85\%$, which rises to $98.6\%$ once the ten predicted quadratic modes are included.

\reply{More directly, the ten- and fifteen-mode quadratic sums have a mutual mismatch of $0.69\%$ at $4\,M_\mathrm{f}$, whereas the smaller truncations change visibly as modes are added. We use this ten-to-fifteen comparison as a convergence test that the omitted couplings involving $(2,2,4)$ no longer change the bundled quadratic waveform considerably at \starttime{}.  These results establish the statement in the main text that the ten-mode bundle is convergent to faithfully represent the quadratic sector.}

\subsection{Linear content of the residual}

\begin{figure}[htbp]
\includegraphics[width=\columnwidth]{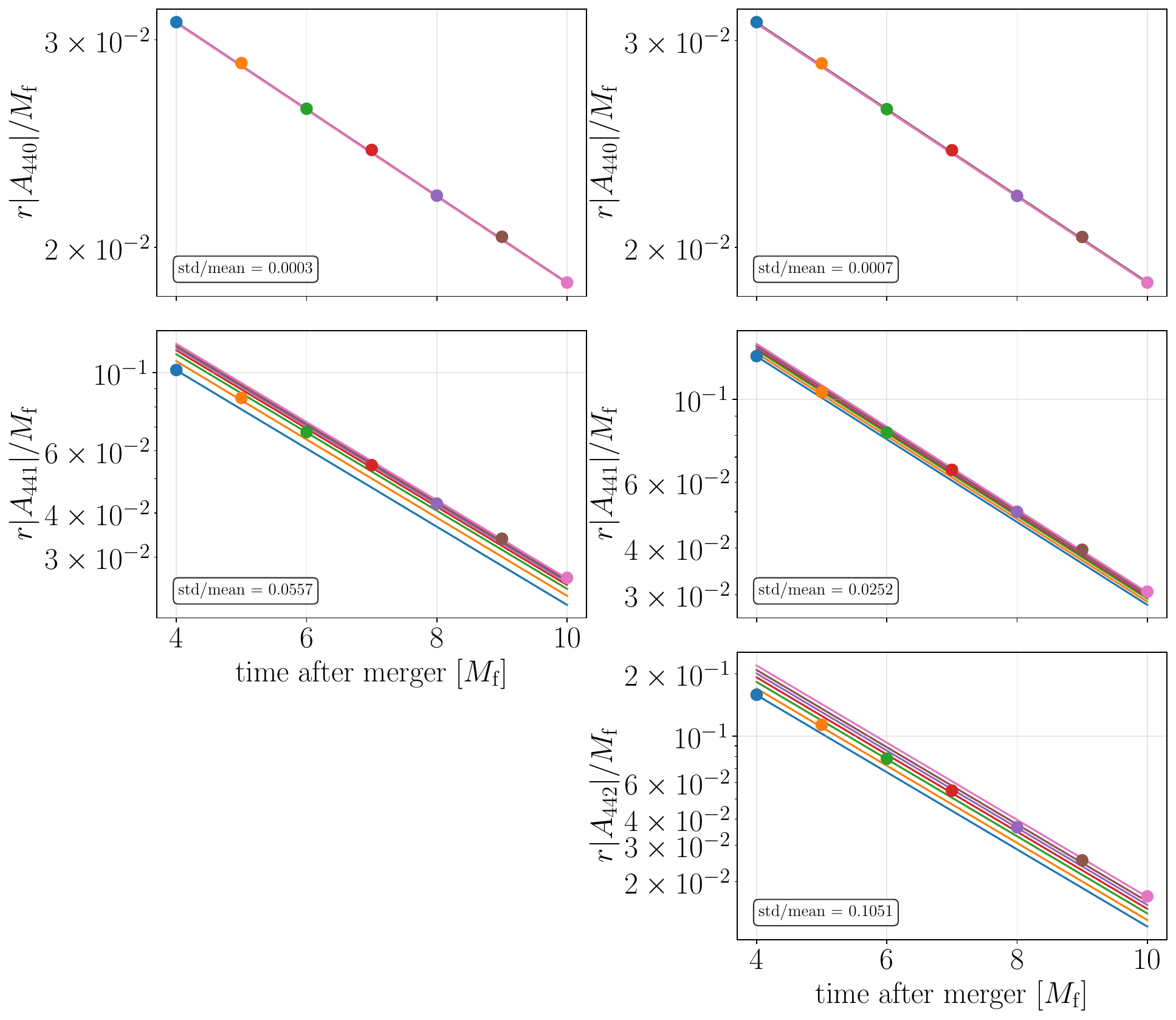}
\caption{
\reply{Dependence of the extracted linear $(4,4,n)$ content on the beginning of the fitting window, after subtracting the ten-mode quadratic prediction from SXS:BBH:3617. The left column shows the $(4,4,0)$ and $(4,4,1)$ amplitudes from a weighted least-squares fit using these two modes, with $(4,4,2)$ setting the error envelope. The right column shows the corresponding three-mode fit including $(4,4,2)$, with $(4,4,3)$ setting the envelope. Dots denote fitting windows beginning at $4$--$10\,M_\mathrm{f}$, curves propagate each result to a common $4\,M_\mathrm{f}$ reference time, and the boxed values give the ratio of the standard deviation to the mean of the propagated amplitudes.}
}
\label{fig:ampconv44}
\end{figure}

\reply{To characterize the residual, we subtract the ten predicted quadratic modes and perform weighted least-squares fits using windows that begin between $4$ and $10\,M_\mathrm{f}$. We compare a $(4,4,0)+(4,4,1)$ fit, whose error envelope is set by $(4,4,2)$, with a three-mode fit that includes $(4,4,2)$ and uses $(4,4,3)$ as the envelope. Propagating all amplitudes to the common $4\,M_\mathrm{f}$ shows the stability across the fitting window choices explicit in \cref{fig:ampconv44}.}
\reply{By contrast, fitting the raw $(4,4)$ multipole without the quadratic modes forces the linear overtones to absorb them and produces a much stronger dependence on where the fitting window begins. The stable QQNM-subtracted fits support the interpretation of the residual as linear $(4,4,n)$ modes.}

\begin{figure}[htbp]
\includegraphics[width=\columnwidth]{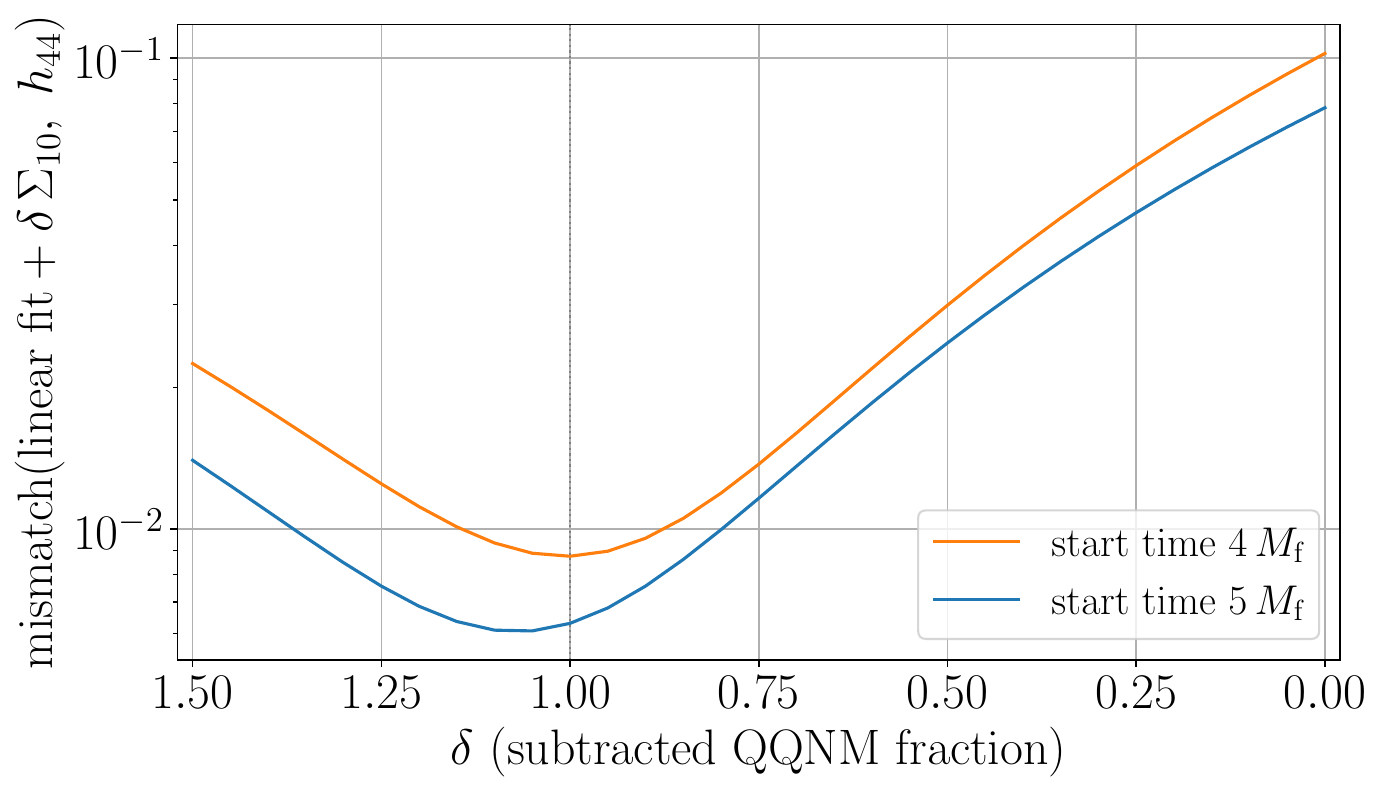}
\caption{
\reply{SXS:BBH:3617 amplitude scan using direct parent-mode fits. For each subtracted fraction $\delta$ of the ten-mode sum $\Sigma_{10}$, an ordinary least-squares $(4,4,0)+(4,4,1)$ fit is performed on $h_{44}-\delta\Sigma_{10}$. The plotted mismatch is evaluated over $[t_0,t_0+15\,M_\mathrm{f}]$ for start times $t_0=4$ and $5\,M_\mathrm{f}$.}
}
\label{fig:linfit-delta}
\end{figure}

\reply{\Cref{fig:linfit-delta} provides a complementary amplitude scan. We subtract a fraction $\delta$ of the predicted ten-mode sum, refit the $(4,4,0)+(4,4,1)$ residual, and evaluate the reconstruction mismatch over $[t_0,t_0+15\,M_\mathrm{f}]$. This window contains the majority of the quadratic mode contents while avoiding late-time numerical contamination. The minimum occurs at $\delta \sim 1.0$ for $t_0=4\,M_\mathrm{f}$ and at $\delta\simeq1.05$ for $t_0=5\,M_\mathrm{f}$. Thus the numerical waveform independently prefers the full quadratic modes to be included.}

We thus conclude that the $(4,4)$ multipole is consistently described as the predicted quadratic bundle plus a linear component starting at \starttime{} after merger.

\reply{We also estimate the possible direct-wave contamination from the results of Ref.~\cite{Oshita:2025qmn}. For a GW150914-like signal the direct wave carries a network \ac{SNR} of $1.6$ from the strain peak, an eighth of the total post-peak \ac{SNR} of $12.8$, and decays at the rate $\kappa \simeq 0.21/M_\mathrm{f}$, roughly twice as fast as the $(2,2,0)$ mode. Assuming the same fractional excitation holds for the $(4,4)$ multipole, whose post-peak \ac{SNR} in GW250114 is $\simeq 3$, the direct wave contributes $\lesssim 0.35$ from the peak, and its fast decay reduces this to $\rho_\mathrm{direct} \lesssim 0.35\, e^{-4 \kappa M_\mathrm{f}} \simeq 0.15$ at the fiducial start time of \starttime{}. This is well below both the quadratic-sum \ac{SNR} ($\sim \expectedsnr$) and unity, the direct wave is therefore negligible for assessing the modal/non-modal content of $(4,4)$.}

\section{Injection studies}
\label{sec:app:injection}

\begin{figure}[t]
\includegraphics[width=\columnwidth]{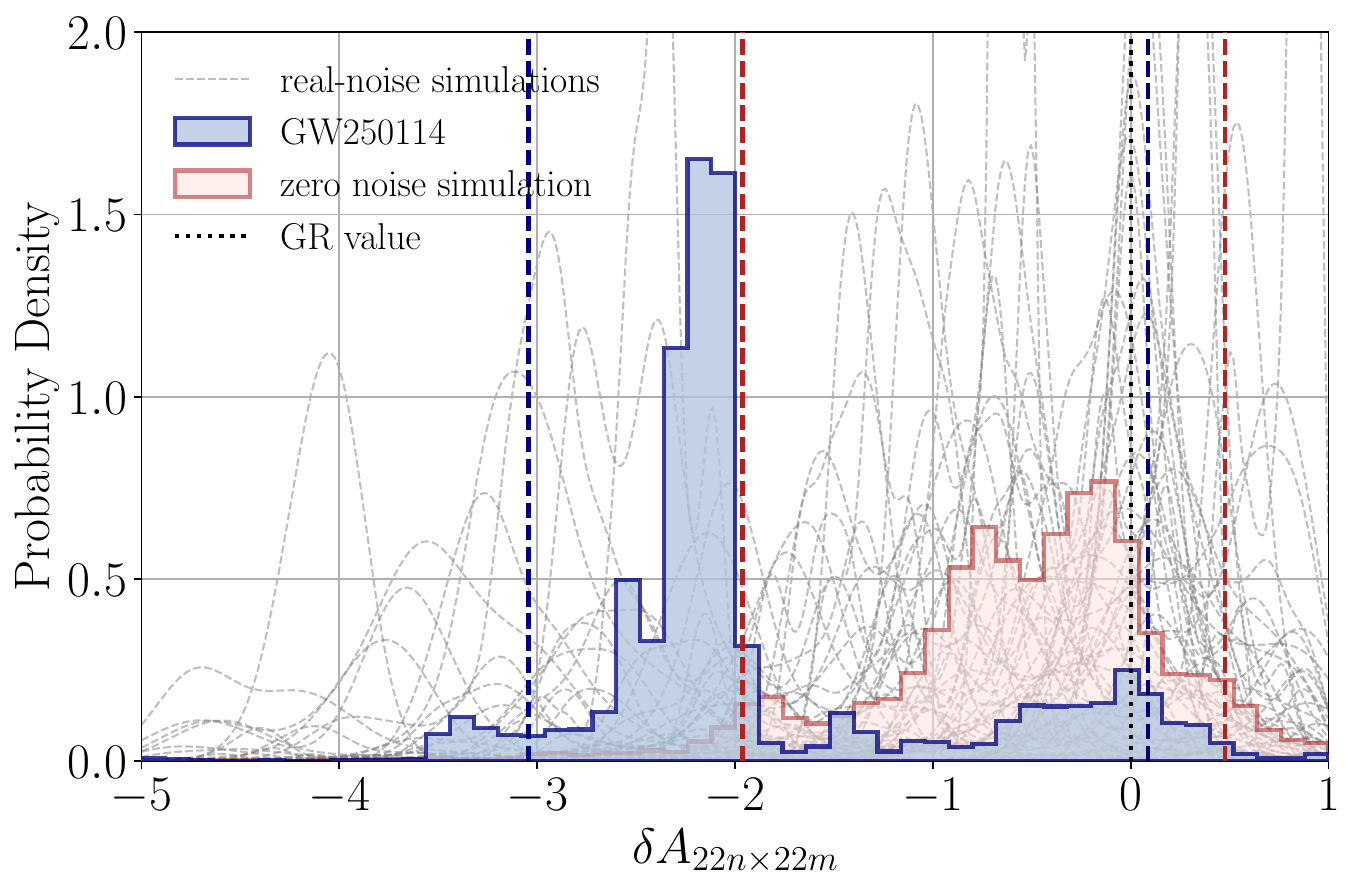}
\caption{\reply{Posterior distributions of the fractional quadratic-mode amplitude deviation $\delta A_{22n\times22m}$, defined by $A_\alpha=(1-\delta A)A^{\rm GR}_\alpha$, where $\alpha$ labels the index of QQNM. The dotted black line marks the \ac{GR} prediction, $\delta A=0$. The filled blue distribution is the GW250114 posterior, with its 90\% credible interval shown by the dashed navy lines. The filled red distribution is the zero-noise \ac{GR} injection, with its 90\% credible interval shown by the dashed red lines. The 50 gray dashed curves are the posteriors obtained after injecting the same maximum likelihood \ac{GR} waveform into different segments of real detector noise.}}
\label{fig:tgr_comparison}
\end{figure}

\reply{The GW250114 posterior favors a quadratic-mode amplitude above the \ac{GR} prediction. Its median is $\delta A=-2.14$, corresponding to $A/A_{\rm GR}=3.14$, and the \ac{GR} value $\delta A=0$ lies at the 7\% posterior quantile. We test whether detector noise can produce a displacement of this magnitude using an injection campaign that resembles the analysis configuration.}

\reply{We inject the maximum-likelihood GW250114 waveform using \texttt{NRSur7dq4}, including the \ac{GR}-predicted quadratic modes, into 50 real-noise segments selected within $[-45,+685]$~s of the merger time. The event itself is excluded. Each injection uses the same exact $6-10~M_\mathrm{f}$ Gaussian parent-amplitude reconstruction, gating and inpainting, priors, and ringdown start time \starttime{} as the data analysis.}

\reply{The real-noise posteriors show substantial scatter (gray curves in \cref{fig:tgr_comparison}). 12 of 50 injections (24\%) have $\delta A=0$ at an equally or more extreme posterior quantile than GW250114. The observed amplitude offset in GW250114 is therefore compatible with the distribution obtained from \ac{GR} injections in real detector noise.}

\reply{We also analyze a zero-noise control simulation using the same noise power spectral density as GW250114 analysis. It gives a median $\delta A=-0.45$ and a 90\% credible interval $[-1.96,0.48]$, the GR-value 0 lies at the 19\% quantile, well within the credible interval (\cref{fig:tgr_comparison}). This control simulation provides no evidence for a systematic offset in the absence of detector noise fluctuation.}

\reply{An independent parameterized effective-one-body analysis of GW250114, using a different waveform family, ringdown treatment, and inference pipeline, found the same mild preference for a $(4,4)$ multipole amplitude above the \ac{GR} prediction \cite{Grimaldi:2026prn}. The analyses share the same detector-noise realization but otherwise use substantially different methods, hence suggesting a noise fluctuation related interpretation of the common offset.}

\bibliography{reference.bib}

@article{Arnaudo:2025uos,
    author = "Arnaudo, Paolo and Carballo, Javier and Withers, Benjamin",
    title = "{Beyond quasinormal modes: A complete mode decomposition of black hole perturbations}",
    eprint = "2510.18956",
    archivePrefix = "arXiv",
    primaryClass = "gr-qc",
    doi = "10.1103/1bcl-q79x",
    journal = "Phys. Rev. D",
    volume = "114",
    number = "2",
    pages = "024025",
    year = "2026"
}

@article{Lagos:2022otp,
    author = "Lagos, Macarena and Hui, Lam",
    title = "{Generation and propagation of nonlinear quasinormal modes of a Schwarzschild black hole}",
    eprint = "2208.07379",
    archivePrefix = "arXiv",
    primaryClass = "gr-qc",
    doi = "10.1103/PhysRevD.107.044040",
    journal = "Phys. Rev. D",
    volume = "107",
    number = "4",
    pages = "044040",
    year = "2023"
}

@article{London:2014cma,
    author = "London, Lionel and Shoemaker, Deirdre and Healy, James",
    title = "{Modeling ringdown: Beyond the fundamental quasinormal modes}",
    eprint = "1404.3197",
    archivePrefix = "arXiv",
    primaryClass = "gr-qc",
    doi = "10.1103/PhysRevD.90.124032",
    journal = "Phys. Rev. D",
    volume = "90",
    number = "12",
    pages = "124032",
    year = "2014",
    note = "[Erratum: Phys.Rev.D 94, 069902 (2016)]"
}

@article{LIGOScientific:2026wpt,
    author = "Abac, A. G. and others",
    collaboration = "LIGO Scientific, VIRGO, KAGRA",
    title = "{GWTC-4.0: Tests of General Relativity. III. Tests of the Remnants}",
    eprint = "2603.19021",
    archivePrefix = "arXiv",
    primaryClass = "gr-qc",
    reportNumber = "LIGO-P2500067",
    month = "3",
    year = "2026",
    journal = "arXiv preprint"
}

@book{jeffreys1998theory,
  title={The theory of probability},
  author={Jeffreys, Harold},
  year={1998},
  publisher={OuP Oxford}
}

@article{Aso:2013eba,
    author = "Aso, Yoichi and Michimura, Yuta and Somiya, Kentaro and Ando, Masaki and Miyakawa, Osamu and Sekiguchi, Takanori and Tatsumi, Daisuke and Yamamoto, Hiroaki",
    collaboration = "KAGRA",
    title = "{Interferometer design of the KAGRA gravitational wave detector}",
    eprint = "1306.6747",
    archivePrefix = "arXiv",
    primaryClass = "gr-qc",
    doi = "10.1103/PhysRevD.88.043007",
    journal = "Phys. Rev. D",
    volume = "88",
    number = "4",
    pages = "043007",
    year = "2013"
}

@article{TheVirgo:2014hva,
      author         = "Acernese, F. and others",
      title          = "{Advanced Virgo: a second-generation interferometric
                        gravitational wave detector}",
      collaboration  = "VIRGO",
      journal        = "Class. Quantum Grav.",
      number         = "2",
      volume         = "32",
      pages          = "024001",
      doi            = "10.1088/0264-9381/32/2/024001",
      year           = "2015",
      eprint         = "1408.3978",
      archivePrefix  = "arXiv",
      primaryClass   = "gr-qc",
      SLACcitation   = "%%CITATION = ARXIV:1408.3978;%%",
}

@article{TheLIGOScientific:2014jea,
    author = "Aasi, J. and others",
    collaboration = "LIGO Scientific",
    title = "{Advanced LIGO}",
    eprint = "1411.4547",
    archivePrefix = "arXiv",
    primaryClass = "gr-qc",
    doi = "10.1088/0264-9381/32/7/074001",
    journal = "Class. Quant. Grav.",
    volume = "32",
    pages = "074001",
    year = "2015"
}

@article{MaganaZertuche:2024ajz,
    author = "Maga{\~n}a Zertuche, Lorena and others",
    title = "{High-precision ringdown surrogate model for nonprecessing binary black holes}",
    eprint = "2408.05300",
    archivePrefix = "arXiv",
    primaryClass = "gr-qc",
    doi = "10.1103/q7sy-g3kl",
    journal = "Phys. Rev. D",
    volume = "112",
    number = "2",
    pages = "024077",
    year = "2025"
}

@article{Pacilio:2024tdl,
    author = "Pacilio, Costantino and Bhagwat, Swetha and Nobili, Francesco and Gerosa, Davide",
    title = "{Flexible mapping of ringdown amplitudes for nonprecessing binary black holes}",
    eprint = "2408.05276",
    archivePrefix = "arXiv",
    primaryClass = "gr-qc",
    doi = "10.1103/PhysRevD.110.103037",
    journal = "Phys. Rev. D",
    volume = "110",
    number = "10",
    pages = "103037",
    year = "2024"
}

@article{LIGOScientific:2019lzm,
    author = "Abbott, Rich and others",
    collaboration = "LIGO Scientific, Virgo",
    title = "{Open data from the first and second observing runs of Advanced LIGO and Advanced Virgo}",
    eprint = "1912.11716",
    archivePrefix = "arXiv",
    primaryClass = "gr-qc",
    reportNumber = "LIGO-P1900206",
    doi = "10.1016/j.softx.2021.100658",
    journal = "SoftwareX",
    volume = "13",
    pages = "100658",
    year = "2021"
}

@article{KAGRA:2023pio,
    author = "Abbott, R. and others",
    collaboration = "KAGRA, VIRGO, LIGO Scientific",
    title = "{Open Data from the Third Observing Run of LIGO, Virgo, KAGRA, and GEO}",
    eprint = "2302.03676",
    archivePrefix = "arXiv",
    primaryClass = "gr-qc",
    reportNumber = "LIGO-P2200316",
    doi = "10.3847/1538-4365/acdc9f",
    journal = "Astrophys. J. Suppl.",
    volume = "267",
    number = "2",
    pages = "29",
    year = "2023"
}

@article{LIGOScientific:2025snk,
    author = "Abac, A. G. and others",
    collaboration = "LIGO Scientific, VIRGO, KAGRA",
    title = "{Open Data from LIGO, Virgo, and KAGRA through the First Part of the Fourth Observing Run}",
    eprint = "2508.18079",
    archivePrefix = "arXiv",
    primaryClass = "gr-qc",
    reportNumber = "LIGO-P2500167",
    month = aug,
    year = "2025",
    journal = "arXiv preprint"
}

@article{Chandra:2025ipu,
    author = "Chandra, Koustav and Calder{\'o}n Bustillo, Juan",
    title = "{Black-hole ringdown analysis with inspiral-merger informed templates and limitations of classical spectroscopy}",
    eprint = "2509.17315",
    archivePrefix = "arXiv",    
    reportNumber = "DCC-P2500566",
    primaryClass = "gr-qc",
    month = sep,
    year = "2025",
    journal = "arXiv preprint"
}

@article{Siegel:2025xgb,
    author = "Siegel, Harrison and Khusid, Nicole M. and Isi, Maximiliano and Farr, Will M.",
    title = "{GW231123 ringdown: interpretation as multimodal Kerr signal}",
    eprint = "2511.02691",
    archivePrefix = "arXiv",
    primaryClass = "gr-qc",
    month = nov,
    year = "2025",
    journal = "arXiv preprint"
}

@article{Zhu:2023fnf,
    author = "Zhu, Hengrui and others",
    title = "{Black hole spectroscopy for precessing binary black hole coalescences}",
    eprint = "2312.08588",
    archivePrefix = "arXiv",
    primaryClass = "gr-qc",
    doi = "10.1103/PhysRevD.111.064052",
    journal = "Phys. Rev. D",
    volume = "111",
    number = "6",
    pages = "064052",
    year = "2025"
}

@article{Scheel:2025jct,
    author = "Scheel, Mark A. and others",
    title = "{The SXS collaboration{\textquoteright}s third catalog of binary black hole simulations}",
    eprint = "2505.13378",
    archivePrefix = "arXiv",
    primaryClass = "gr-qc",
    doi = "10.1088/1361-6382/adfd34",
    journal = "Class. Quant. Grav.",
    volume = "42",
    number = "19",
    pages = "195017",
    year = "2025"
}

@article{Giesler:2024hcr,
    author = "Giesler, Matthew and others",
    title = "{Overtones and nonlinearities in binary black hole ringdowns}",
    eprint = "2411.11269",
    archivePrefix = "arXiv",
    primaryClass = "gr-qc",
    reportNumber = "YITP-24-154, RIKEN-iTHEMS-Report-24",
    doi = "10.1103/PhysRevD.111.084041",
    journal = "Phys. Rev. D",
    volume = "111",
    number = "8",
    pages = "084041",
    year = "2025"
}

@article{Dong:2025igh,
    author = "Dong, Yiming and Wang, Ziming and Wang, Hai-Tian and Zhao, Junjie and Shao, Lijing",
    title = "{A practical Bayesian method for gravitational-wave ringdown analysis with multiple modes}",
    eprint = "2502.01093",
    archivePrefix = "arXiv",
    primaryClass = "gr-qc",
    journal = "arXiv preprint",
    year = "2025"
}

@article{Wang:2024jlz,
    author = "Wang, Hai-Tian and Yim, Garvin and Chen, Xian and Shao, Lijing",
    title = "{Gravitational Wave Ringdown Analysis Using the F -statistic}",
    eprint = "2409.00970",
    archivePrefix = "arXiv",
    primaryClass = "gr-qc",
    doi = "10.3847/1538-4357/ad7096",
    journal = "Astrophys. J.",
    volume = "974",
    number = "2",
    pages = "230",
    year = "2024"
}

@article{LIGOScientific:2021sio,
    author = "Abbott, R. and others",
    collaboration = "LIGO Scientific, VIRGO, KAGRA",
    title = "{Tests of General Relativity with GWTC-3}",
    eprint = "2112.06861",
    archivePrefix = "arXiv",
    primaryClass = "gr-qc",
    reportNumber = "LIGO-P2100275",
    doi = "10.1103/PhysRevD.112.084080",
    journal = "Phys. Rev. D",
    volume = "112",
    number = "8",
    pages = "084080",
    year = "2025"
}

@article{LIGOScientific:2020tif,
    author = "Abbott, R. and others",
    collaboration = "LIGO Scientific, Virgo",
    title = "{Tests of general relativity with binary black holes from the second LIGO-Virgo gravitational-wave transient catalog}",
    eprint = "2010.14529",
    archivePrefix = "arXiv",
    primaryClass = "gr-qc",
    reportNumber = "LIGO-P2000091",
    doi = "10.1103/PhysRevD.103.122002",
    journal = "Phys. Rev. D",
    volume = "103",
    number = "12",
    pages = "122002",
    year = "2021"
}

@article{LIGOScientific:2019fpa,
    author = "Abbott, B. P. and others",
    collaboration = "LIGO Scientific, Virgo",
    title = "{Tests of General Relativity with the Binary Black Hole Signals from the LIGO-Virgo Catalog GWTC-1}",
    eprint = "1903.04467",
    archivePrefix = "arXiv",
    primaryClass = "gr-qc",
    reportNumber = "LIGO-P1800316",
    doi = "10.1103/PhysRevD.100.104036",
    journal = "Phys. Rev. D",
    volume = "100",
    number = "10",
    pages = "104036",
    year = "2019"
}

@article{KAGRA:2025oiz,
    author = "Abac, A. G. and others",
    collaboration = "KAGRA, Virgo, LIGO Scientific",
    title = "{GW250114: Testing Hawking{\textquoteright}s Area Law and the Kerr Nature of Black Holes}",
    eprint = "2509.08054",
    archivePrefix = "arXiv",
    primaryClass = "gr-qc",
    reportNumber = "LIGO-P2500421",
    doi = "10.1103/kw5g-d732",
    journal = "Phys. Rev. Lett.",
    volume = "135",
    number = "11",
    pages = "111403",
    year = "2025"
}

@article{Ma:2024qcv,
    author = "Ma, Sizheng and Yang, Huan",
    title = "{Excitation of quadratic quasinormal modes for Kerr black holes}",
    eprint = "2401.15516",
    archivePrefix = "arXiv",
    primaryClass = "gr-qc",
    doi = "10.1103/PhysRevD.109.104070",
    journal = "Phys. Rev. D",
    volume = "109",
    number = "10",
    pages = "104070",
    year = "2024"
}

@article{Khera:2023oyf,
    author = "Khera, Neev and Ribes Metidieri, Ariadna and Bonga, B{\'e}atrice and Jim{\'e}nez Forteza, Xisco and Krishnan, Badri and Poisson, Eric and Pook-Kolb, Daniel and Schnetter, Erik and Yang, Huan",
    title = "{Nonlinear Ringdown at the Black Hole Horizon}",
    eprint = "2306.11142",
    archivePrefix = "arXiv",
    primaryClass = "gr-qc",
    doi = "10.1103/PhysRevLett.131.231401",
    journal = "Phys. Rev. Lett.",
    volume = "131",
    number = "23",
    pages = "231401",
    year = "2023"
}

@article{Berti:2025hly,
    author = "Berti, Emanuele and others",
    title = "{Black hole spectroscopy: from theory to experiment}",
    eprint = "2505.23895",
    archivePrefix = "arXiv",
    primaryClass = "gr-qc",
    month = "5",
    year = "2025",
    journal = ""
}

@article{LIGOScientific:2025obp,
    collaboration = "LIGO Scientific, VIRGO, KAGRA",
    title = "{Black Hole Spectroscopy and Tests of General Relativity with GW250114}",
    eprint = "2509.08099",
    archivePrefix = "arXiv",
    primaryClass = "gr-qc",
    reportNumber = "LIGO P2500461",
    month = "9",
    year = "2025",
    journal = ""
}

@article{Lu:2025vol,
    author = "Lu, Neil and Ma, Sizheng and Piccinni, Ornella J. and Chen, Yanbei and Sun, Ling",
    title = "{GW250114 reveals signatures of post-merger black-hole horizon}",
    eprint = "2510.01001",
    archivePrefix = "arXiv",
    primaryClass = "gr-qc",
    doi = "10.1038/s41586-026-10696-0",
    journal = "Nature",
    volume = "655",
    number = "8121",
    pages = "71--75",
    year = "2026"
}

@article{Siegel:2023lxl,
    author = "Siegel, Harrison and Isi, Maximiliano and Farr, Will M.",
    title = "{Ringdown of GW190521: Hints of multiple quasinormal modes with a precessional interpretation}",
    eprint = "2307.11975",
    archivePrefix = "arXiv",
    primaryClass = "gr-qc",
    reportNumber = "LIGO-P2300214",
    doi = "10.1103/PhysRevD.108.064008",
    journal = "Phys. Rev. D",
    volume = "108",
    number = "6",
    pages = "064008",
    year = "2023"
}

@article{Capano:2021etf,
    author = "Capano, Collin D. and Cabero, Miriam and Westerweck, Julian and Abedi, Jahed and Kastha, Shilpa and Nitz, Alexander H. and Wang, Yi-Fan and Nielsen, Alex B. and Krishnan, Badri",
    title = "{Multimode Quasinormal Spectrum from a Perturbed Black Hole}",
    eprint = "2105.05238",
    archivePrefix = "arXiv",
    primaryClass = "gr-qc",
    doi = "10.1103/PhysRevLett.131.221402",
    journal = "Phys. Rev. Lett.",
    volume = "131",
    number = "22",
    pages = "221402",
    year = "2023"
}

@article{Correia:2023bfn,
    author = "Correia, Alex and Wang, Yi-Fan and Westerweck, Julian and Capano, Collin D.",
    title = "{Low evidence for ringdown overtone in GW150914 when marginalizing over time and sky location uncertainty}",
    eprint = "2312.14118",
    archivePrefix = "arXiv",
    primaryClass = "gr-qc",
    doi = "10.1103/PhysRevD.110.L041501",
    journal = "Phys. Rev. D",
    volume = "110",
    number = "4",
    pages = "L041501",
    year = "2024"
}

@article{Dreyer:2003bv,
    author = "Dreyer, Olaf and Kelly, Bernard J. and Krishnan, Badri and Finn, Lee Samuel and Garrison, David and Lopez-Aleman, Ramon",
    title = "{Black hole spectroscopy: Testing general relativity through gravitational wave observations}",
    eprint = "gr-qc/0309007",
    archivePrefix = "arXiv",
    doi = "10.1088/0264-9381/21/4/003",
    journal = "Class. Quant. Grav.",
    volume = "21",
    pages = "787--804",
    year = "2004"
}

@article{qnm1,
	Author = {Vishveshwara, C.  V. },
	Da = {1970/08/01},
	Doi = {10.1038/227936a0},
	Id = {VISHVESHWARA1970},
	Isbn = {1476-4687},
	Journal = {Nature},
	Number = {5261},
	Pages = {936--938},
	Title = {Scattering of Gravitational Radiation by a Schwarzschild Black-hole},
	Ty = {JOUR},
	Url = {https://doi.org/10.1038/227936a0},
	Volume = {227},
	Year = {1970}}

@article{Nohairseminal3,
	Author = {Carter, B.},
	Doi = {10.1103/PhysRevLett.26.331},
	Issue = {6},
	Journal = {Phys. Rev. Lett.},
	Month = {Feb},
	Numpages = {0},
	Pages = {331--333},
	Publisher = {American Physical Society},
	Title = {Axisymmetric Black Hole Has Only Two Degrees of Freedom},
	Url = {https://link.aps.org/doi/10.1103/PhysRevLett.26.331},
	Volume = {26},
	Year = {1971}}

@article{Nohairseminal2,
	Author = {Israel, Werner},
	Day = {01},
	Doi = {10.1007/BF01645859},
	Issn = {1432-0916},
	Journal = {Communications in Mathematical Physics},
	Month = {Sep},
	Number = {3},
	Pages = {245--260},
	Title = {Event horizons in static electrovac space-times},
	Url = {https://doi.org/10.1007/BF01645859},
	Volume = {8},
	Year = {1968}}

@article{Nohairseminal1,
	Author = {Israel, Werner},
	Doi = {10.1103/PhysRev.164.1776},
	Issue = {5},
	Journal = {Phys. Rev.},
	Month = {Dec},
	Numpages = {0},
	Pages = {1776--1779},
	Publisher = {American Physical Society},
	Title = {Event Horizons in Static Vacuum Space-Times},
	Url = {https://link.aps.org/doi/10.1103/PhysRev.164.1776},
	Volume = {164},
	Year = {1967}}

@article{LIGOScientific:2025rsn,
    author = "Abac, A. G. and others",
    collaboration = "LIGO Scientific, VIRGO, KAGRA",
    title = "{GW231123: a Binary Black Hole Merger with Total Mass 190-265 $M_{\odot}$}",
    eprint = "2507.08219",
    archivePrefix = "arXiv",
    primaryClass = "astro-ph.HE",
    reportNumber = "DCC: P2500026-v6",
    month = "7",
    year = "2025",
    journal = ""
}

@article{LIGOScientific:2020iuh,
    author = "Abbott, R. and others",
    collaboration = "LIGO Scientific, Virgo",
    title = "{GW190521: A Binary Black Hole Merger with a Total Mass of $150  M_{\odot}$}",
    eprint = "2009.01075",
    archivePrefix = "arXiv",
    primaryClass = "gr-qc",
    doi = "10.1103/PhysRevLett.125.101102",
    journal = "Phys. Rev. Lett.",
    volume = "125",
    number = "10",
    pages = "101102",
    year = "2020"
}

@article{LIGOScientific:2020ufj,
    author = "Abbott, R. and others",
    collaboration = "LIGO Scientific, Virgo",
    title = "{Properties and Astrophysical Implications of the 150 M$_\odot$ Binary Black Hole Merger GW190521}",
    eprint = "2009.01190",
    archivePrefix = "arXiv",
    primaryClass = "astro-ph.HE",
    reportNumber = "LIGO-P2000021",
    doi = "10.3847/2041-8213/aba493",
    journal = "Astrophys. J. Lett.",
    volume = "900",
    number = "1",
    pages = "L13",
    year = "2020"
}

@article{Chavda:2024awq,
    author = "Chavda, Ameya and Lagos, Macarena and Hui, Lam",
    title = "{The impact of initial conditions on quasi-normal modes}",
    eprint = "2412.03435",
    archivePrefix = "arXiv",
    primaryClass = "gr-qc",
    doi = "10.1088/1475-7516/2025/07/084",
    journal = "JCAP",
    volume = "07",
    pages = "084",
    year = "2025"
}

@article{DeAmicis:2025xuh,
    author = "De Amicis, Marina and Cannizzaro, Enrico and Carullo, Gregorio and Sberna, Laura",
    title = "{Dynamical quasinormal mode excitation}",
    eprint = "2506.21668",
    archivePrefix = "arXiv",
    primaryClass = "gr-qc",
    month = jun,
    year = "2025",
    journal = "arXiv preprint"
}

@article{Carullo:2019flw,
    author = "Carullo, Gregorio and Del Pozzo, Walter and Veitch, John",
    title = "{Observational Black Hole Spectroscopy: A time-domain multimode analysis of GW150914}",
    eprint = "1902.07527",
    archivePrefix = "arXiv",
    primaryClass = "gr-qc",
    doi = "10.1103/PhysRevD.99.123029",
    journal = "Phys. Rev. D",
    volume = "99",
    number = "12",
    pages = "123029",
    year = "2019",
    note = "[Erratum: Phys.Rev.D 100, 089903 (2019)]"
}

@article{Wang:2023ljx,
    author = "Wang, Yi-Fan and Capano, Collin D. and Abedi, Jahed and Kastha, Shilpa and Krishnan, Badri and Nielsen, Alex B. and Nitz, Alexander H. and Westerweck, Julian",
    title = "{Gating-and-inpainting perspective on GW150914 ringdown overtone: Understanding the data analysis systematics}",
    eprint = "2310.19645",
    archivePrefix = "arXiv",
    primaryClass = "gr-qc",
    reportNumber = "LIGO-P2300340",
    doi = "10.1103/3gqn-297f",
    journal = "Phys. Rev. D",
    volume = "112",
    number = "8",
    pages = "083023",
    year = "2025"
}

@article{Finch:2021qph,
    author = "Finch, Eliot and Moore, Christopher J.",
    title = "{Frequency-domain analysis of black-hole ringdowns}",
    eprint = "2108.09344",
    archivePrefix = "arXiv",
    primaryClass = "gr-qc",
    reportNumber = "LIGO document number P2100303-v2",
    doi = "10.1103/PhysRevD.104.123034",
    journal = "Phys. Rev. D",
    volume = "104",
    number = "12",
    pages = "123034",
    year = "2021"
}

@article{Ma:2022wpv,
    author = "Ma, Sizheng and Mitman, Keefe and Sun, Ling and Deppe, Nils and H{\'e}bert, Fran{\c{c}}ois and Kidder, Lawrence E. and Moxon, Jordan and Throwe, William and Vu, Nils L. and Chen, Yanbei",
    title = "{Quasinormal-mode filters: A new approach to analyze the gravitational-wave ringdown of binary black-hole mergers}",
    eprint = "2207.10870",
    archivePrefix = "arXiv",
    primaryClass = "gr-qc",
    doi = "10.1103/PhysRevD.106.084036",
    journal = "Phys. Rev. D",
    volume = "106",
    number = "8",
    pages = "084036",
    year = "2022"
}

@article{Varma:2019csw,
    author = "Varma, Vijay and Field, Scott E. and Scheel, Mark A. and Blackman, Jonathan and Gerosa, Davide and Stein, Leo C. and Kidder, Lawrence E. and Pfeiffer, Harald P.",
    title = "{Surrogate models for precessing binary black hole simulations with unequal masses}",
    eprint = "1905.09300",
    archivePrefix = "arXiv",
    primaryClass = "gr-qc",
    doi = "10.1103/PhysRevResearch.1.033015",
    journal = "Phys. Rev. Research.",
    volume = "1",
    pages = "033015",
    year = "2019"
}

@article{Ma:2023vvr,
    author = "Ma, Sizheng and Sun, Ling and Chen, Yanbei",
    title = "{Using rational filters to uncover the first ringdown overtone in GW150914}",
    eprint = "2301.06639",
    archivePrefix = "arXiv",
    primaryClass = "gr-qc",
    doi = "10.1103/PhysRevD.107.084010",
    journal = "Phys. Rev. D",
    volume = "107",
    number = "8",
    pages = "084010",
    year = "2023"
}

@article{Ma:2023cwe,
    author = "Ma, Sizheng and Sun, Ling and Chen, Yanbei",
    title = "{Black Hole Spectroscopy by Mode Cleaning}",
    eprint = "2301.06705",
    archivePrefix = "arXiv",
    primaryClass = "gr-qc",
    doi = "10.1103/PhysRevLett.130.141401",
    journal = "Phys. Rev. Lett.",
    volume = "130",
    number = "14",
    pages = "141401",
    year = "2023"
}

@article{Oshita:2025qmn,
    author = "Oshita, Naritaka and Ma, Sizheng and Chen, Yanbei and Yang, Huan",
    title = "{Probing Direct Waves in Black Hole Ringdowns}",
    eprint = "2509.09165",
    archivePrefix = "arXiv",
    primaryClass = "gr-qc",
    reportNumber = "YITP-25-141, RIKEN-iTHEMS-Report-25",
    month = "9",
    year = "2025",
    journal = ""
}

@article{Isi:2021iql,
    author = "Isi, Maximiliano and Farr, Will M.",
    title = "{Analyzing black-hole ringdowns}",
    eprint = "2107.05609",
    archivePrefix = "arXiv",
    primaryClass = "gr-qc",
    reportNumber = "LIGO-P2100227",
    month = "7",
    year = "2021",
    journal = ""
}

@article{Ma:2024hzq,
    author = "Ma, Sizheng and Scheel, Mark A. and Moxon, Jordan and Nelli, Kyle C. and Deppe, Nils and Kidder, Lawrence E. and Throwe, William and Vu, Nils L.",
    title = "{Merging black holes with Cauchy-characteristic matching: Computation of late-time tails}",
    eprint = "2412.06906",
    archivePrefix = "arXiv",
    primaryClass = "gr-qc",
    doi = "10.1103/jd26-8q5w",
    journal = "Phys. Rev. D",
    volume = "112",
    number = "2",
    pages = "024003",
    year = "2025"
}

@article{DeAmicis:2024eoy,
    author = "De Amicis, Marina and others",
    title = "{Late-time tails in nonlinear evolutions of merging black holes}",
    eprint = "2412.06887",
    archivePrefix = "arXiv",
    primaryClass = "gr-qc",
    month = "12",
    year = "2024",
    journal = "",
}

@article{Islam:2024vro,
    author = "Islam, Tousif and Faggioli, Guglielmo and Khanna, Gaurav and Field, Scott E. and van de Meent, Maarten and Buonanno, Alessandra",
    title = "{Phenomenology and origin of late-time tails in eccentric binary black hole mergers}",
    eprint = "2407.04682",
    archivePrefix = "arXiv",
    primaryClass = "gr-qc",
    doi = "10.1103/191t-5svc",
    journal = "Phys. Rev. D",
    volume = "112",
    number = "2",
    pages = "024061",
    year = "2025"
}

@article{DeAmicis:2024not,
    author = "De Amicis, Marina and Albanesi, Simone and Carullo, Gregorio",
    title = "{Inspiral-inherited ringdown tails}",
    eprint = "2406.17018",
    archivePrefix = "arXiv",
    primaryClass = "gr-qc",
    doi = "10.1103/PhysRevD.110.104005",
    journal = "Phys. Rev. D",
    volume = "110",
    number = "10",
    pages = "104005",
    year = "2024"
}

@article{Khera:2024bjs,
    author = "Khera, Neev and Ma, Sizheng and Yang, Huan",
    title = "{Quadratic Mode Couplings in Rotating Black Holes and Their Detectability}",
    eprint = "2410.14529",
    archivePrefix = "arXiv",
    primaryClass = "gr-qc",
    doi = "10.1103/PhysRevLett.134.211404",
    journal = "Phys. Rev. Lett.",
    volume = "134",
    number = "21",
    pages = "211404",
    year = "2025"
}

@article{Mitman:2022qdl,
    author = "Mitman, Keefe and others",
    title = "{Nonlinearities in Black Hole Ringdowns}",
    eprint = "2208.07380",
    archivePrefix = "arXiv",
    primaryClass = "gr-qc",
    doi = "10.1103/PhysRevLett.130.081402",
    journal = "Phys. Rev. Lett.",
    volume = "130",
    number = "8",
    pages = "081402",
    year = "2023"
}

@article{Cheung:2022rbm,
    author = "Cheung, Mark Ho-Yeuk and others",
    title = "{Nonlinear Effects in Black Hole Ringdown}",
    eprint = "2208.07374",
    archivePrefix = "arXiv",
    primaryClass = "gr-qc",
    doi = "10.1103/PhysRevLett.130.081401",
    journal = "Phys. Rev. Lett.",
    volume = "130",
    number = "8",
    pages = "081401",
    year = "2023"
}

@article{Isi:2019aib,
    author = "Isi, Maximiliano and Giesler, Matthew and Farr, Will M. and Scheel, Mark A. and Teukolsky, Saul A.",
    title = "{Testing the no-hair theorem with GW150914}",
    eprint = "1905.00869",
    archivePrefix = "arXiv",
    primaryClass = "gr-qc",
    reportNumber = "LIGO-P1900135",
    doi = "10.1103/PhysRevLett.123.111102",
    journal = "Phys. Rev. Lett.",
    volume = "123",
    number = "11",
    pages = "111102",
    year = "2019"
}

@article{LIGOScientific:2016lio,
    author = "Abbott, B. P. and others",
    collaboration = "LIGO Scientific, Virgo",
    title = "{Tests of general relativity with GW150914}",
    eprint = "1602.03841",
    archivePrefix = "arXiv",
    primaryClass = "gr-qc",
    reportNumber = "LIGO-P1500213",
    doi = "10.1103/PhysRevLett.116.221101",
    journal = "Phys. Rev. Lett.",
    volume = "116",
    number = "22",
    pages = "221101",
    year = "2016",
    note = "[Erratum: Phys.Rev.Lett. 121, 129902 (2018)]"
}

@article{Brito:2018rfr,
    author = "Brito, Richard and Buonanno, Alessandra and Raymond, Vivien",
    title = "{Black-hole Spectroscopy by Making Full Use of Gravitational-Wave Modeling}",
    eprint = "1805.00293",
    archivePrefix = "arXiv",
    primaryClass = "gr-qc",
    doi = "10.1103/PhysRevD.98.084038",
    journal = "Phys. Rev. D",
    volume = "98",
    number = "8",
    pages = "084038",
    year = "2018"
}

@article{Ghosh:2021mrv,
    author = "Ghosh, Abhirup and Brito, Richard and Buonanno, Alessandra",
    title = "{Constraints on quasinormal-mode frequencies with LIGO-Virgo binary{\textendash}black-hole observations}",
    eprint = "2104.01906",
    archivePrefix = "arXiv",
    primaryClass = "gr-qc",
    reportNumber = "LIGO-P2100106",
    doi = "10.1103/PhysRevD.103.124041",
    journal = "Phys. Rev. D",
    volume = "103",
    number = "12",
    pages = "124041",
    year = "2021"
}

@article{Maggio:2022hre,
    author = "Maggio, Elisa and Silva, Hector O. and Buonanno, Alessandra and Ghosh, Abhirup",
    title = "{Tests of general relativity in the nonlinear regime: A parametrized plunge-merger-ringdown gravitational waveform model}",
    eprint = "2212.09655",
    archivePrefix = "arXiv",
    primaryClass = "gr-qc",
    reportNumber = "LIGO-P2200377",
    doi = "10.1103/PhysRevD.108.024043",
    journal = "Phys. Rev. D",
    volume = "108",
    number = "2",
    pages = "024043",
    year = "2023"
}

@article{Pompili:2025cdc,
    author = "Pompili, Lorenzo and Maggio, Elisa and Silva, Hector O. and Buonanno, Alessandra",
    title = "{Parametrized spin-precessing inspiral-merger-ringdown waveform model for tests of general relativity}",
    eprint = "2504.10130",
    archivePrefix = "arXiv",
    primaryClass = "gr-qc",
    doi = "10.1103/ng8w-98sz",
    journal = "Phys. Rev. D",
    volume = "111",
    number = "12",
    pages = "124040",
    year = "2025"
}

@article{Grimaldi:2026prn,
    author = "Grimaldi, Leonardo and Maggio, Elisa and Pompili, Lorenzo and Buonanno, Alessandra",
    title = "{Plunge-merger-ringdown tests of general relativity with GW250114}",
    eprint = "2601.13173",
    archivePrefix = "arXiv",
    primaryClass = "gr-qc",
    doi = "10.1103/g19c-734l",
    journal = "Phys. Rev. D",
    volume = "113",
    number = "6",
    pages = "L061506",
    year = "2026"
}

@article{Biwer_2019,
   title={PyCBC Inference: A Python-based Parameter Estimation Toolkit for Compact Binary Coalescence Signals},
   volume={131},
   ISSN={1538-3873},
   url={http://dx.doi.org/10.1088/1538-3873/aaef0b},
   DOI={10.1088/1538-3873/aaef0b},
   number={996},
   journal={Publications of the Astronomical Society of the Pacific},
   publisher={IOP Publishing},
   author={Biwer, C. M. and Capano, Collin D. and De, Soumi and Cabero, Miriam and Brown, Duncan A. and Nitz, Alexander H. and Raymond, V.},
   year={2019},
   month={Jan},
   pages={024503}
}

@article{speagle:2019,
    author = {Speagle, Joshua S},
    title = "{dynesty: a dynamic nested sampling package for estimating Bayesian posteriors and evidences}",
    journal = {Monthly Notices of the Royal Astronomical Society},
    volume = {493},
    number = {3},
    pages = {3132-3158},
    year = {2020},
    month = {02},
    issn = {0035-8711},
    doi = {10.1093/mnras/staa278},
    url = {https://doi.org/10.1093/mnras/staa278},
    eprint = {https://academic.oup.com/mnras/article-pdf/493/3/3132/32890730/staa278.pdf},
}

@article{Correia:2023ipz,
    author = "Correia, Alex and Capano, Collin D.",
    title = "{Sky marginalization in black hole spectroscopy and tests of the area theorem}",
    eprint = "2312.15146",
    archivePrefix = "arXiv",
    primaryClass = "gr-qc",
    doi = "10.1103/PhysRevD.110.044018",
    journal = "Phys. Rev. D",
    volume = "110",
    number = "4",
    pages = "044018",
    year = "2024"
}

@article{Yang:2017zxs,
    author = "Yang, Huan and Yagi, Kent and Blackman, Jonathan and Lehner, Luis and Paschalidis, Vasileios and Pretorius, Frans and Yunes, Nicol{\'a}s",
    title = "{Black hole spectroscopy with coherent mode stacking}",
    eprint = "1701.05808",
    archivePrefix = "arXiv",
    primaryClass = "gr-qc",
    doi = "10.1103/PhysRevLett.118.161101",
    journal = "Phys. Rev. Lett.",
    volume = "118",
    number = "16",
    pages = "161101",
    year = "2017"
}

@article{Hughes:2019zmt,
    author = "Hughes, Scott A. and Apte, Anuj and Khanna, Gaurav and Lim, Halston",
    title = "{Learning about black hole binaries from their ringdown spectra}",
    eprint = "1901.05900",
    archivePrefix = "arXiv",
    primaryClass = "gr-qc",
    doi = "10.1103/PhysRevLett.123.161101",
    journal = "Phys. Rev. Lett.",
    volume = "123",
    number = "16",
    pages = "161101",
    year = "2019"
}

@article{Lim:2022veo,
    author = "Lim, Halston and Hughes, Scott A. and Khanna, Gaurav",
    title = "{Measuring quasinormal mode amplitudes with misaligned binary black hole ringdowns}",
    eprint = "2204.06007",
    archivePrefix = "arXiv",
    primaryClass = "gr-qc",
    doi = "10.1103/PhysRevD.105.124030",
    journal = "Phys. Rev. D",
    volume = "105",
    number = "12",
    pages = "124030",
    year = "2022"
}

@article{Zackay:2019kkv,
    author = "Zackay, Barak and Venumadhav, Tejaswi and Roulet, Javier and Dai, Liang and Zaldarriaga, Matias",
    title = "{Detecting gravitational waves in data with non-stationary and non-Gaussian noise}",
    eprint = "1908.05644",
    archivePrefix = "arXiv",
    primaryClass = "astro-ph.IM",
    doi = "10.1103/PhysRevD.104.063034",
    journal = "Phys. Rev. D",
    volume = "104",
    number = "6",
    pages = "063034",
    year = "2021"
}

@article{Ripley:2022ypi,
    author = "Ripley, Justin L.",
    title = "{Computing the quasinormal modes and eigenfunctions for the Teukolsky equation using horizon penetrating, hyperboloidally compactified coordinates}",
    eprint = "2202.03837",
    archivePrefix = "arXiv",
    primaryClass = "gr-qc",
    doi = "10.1088/1361-6382/ac776d",
    journal = "Class. Quant. Grav.",
    volume = "39",
    number = "14",
    pages = "145009",
    year = "2022"
}

@article{Yi:2024elj,
    author = "Yi, Sophia and Kuntz, Adrien and Barausse, Enrico and Berti, Emanuele and Cheung, Mark Ho-Yeuk and Kritos, Konstantinos and Maselli, Andrea",
    title = "{Nonlinear quasinormal mode detectability with next-generation gravitational wave detectors}",
    eprint = "2403.09767",
    archivePrefix = "arXiv",
    primaryClass = "gr-qc",
    reportNumber = "ET-0081A-24",
    doi = "10.1103/PhysRevD.109.124029",
    journal = "Phys. Rev. D",
    volume = "109",
    number = "12",
    pages = "124029",
    year = "2024"
}

\end{document}